\documentclass[structabstract]{aa}  
\usepackage{graphicx}

\usepackage{epstopdf}
\usepackage{natbib}
\bibpunct{(}{)}{;}{a}{}{,} 

\usepackage{txfonts}
\def\4he{$^4$He}

\def\kms{\mathrm{km\,s}^{-1}}
\def\e#1{\times 10^{#1}}

\def\msol{M_\odot}
\def\lsol{L_\odot}
\def\sec#1{Sect. \ref{#1}}
\def\eq#1{Eq. \ref{#1}}

\def\up#1{$^{#1}$}
\def\down#1{$_{#1}$}

\def\h2{$\mathrm{H}_2$}
\def\nh3{$\mathrm{NH}_3$}
\def\mic{\mathrm{ }\mu\mathrm{m}}
\def\spy{\;\msol \;\mathrm{yr}^{-1}}
\def\um{\;\mu\mathrm{m}}

\def\citeapos#1{\citeauthor{#1}'s (\citeyear{#1})}  

\begin{document}
   \title{Detailed modelling of the circumstellar molecular line emission of the S-type AGB star W Aquilae\thanks{{\it Herschel} is an ESA space observatory with science instruments provided by European-led Principal Investigator consortia and with important participation from NASA.}}
   
   \titlerunning{Detailed modelling of the S-type AGB star W Aquilae}

   \author{T. Danilovich
          \inst{1}
          \and
          P. Bergman \inst{1}
          \and
          K. Justtanont \inst{1}
          \and
          R. Lombaert \inst{4}
          \and
          M. Maercker \inst{1,2}
          \and
          H. Olofsson \inst{1}
          \and
          S. Ramstedt \inst{3}
          \and 
          P. Royer \inst{4}
          }

   \institute{Onsala Space Observatory, Department of Earth and Space Sciences, Chalmers University of Technology,
              SE-439 92 Onsala, Sweden
              \and 
              Argelander Institute for Astronomy, University of Bonn, Auf dem H\"ugel 71, 53121 Bonn, Germany
              \and
              Department of Physics and Astronomy, Uppsala University, Box 515, 751 20 Uppsala, Sweden  
              \and
              Instituut voor Sterrenkunde, KU Leuven, Celestijnenlaan 200D, 3001 Leuven, Belgium
             \\ \email{taissa@chalmers.se}}

   \date{Received 10 October 2013; accepted 7 August 2014}

 
  \abstract
   {S-type AGB stars have a C/O ratio which suggests that they are transition objects between oxygen-rich M-type stars and carbon-rich C-type stars. As such, their circumstellar compositions of gas and dust are thought to be sensitive to their precise C/O ratio, and it is therefore of particular interest to examine their circumstellar properties.}
   {We present new Herschel HIFI {and PACS} sub-millimetre and far-infrared line observations of several molecular species towards the S-type AGB  star W~Aql. We use these observations, which probe a wide range of gas temperatures, to constrain the circumstellar properties of W~Aql, including mass-loss rate 
and molecular abundances.}
   {We used radiative transfer codes to model the circumstellar dust and molecular line emission to determine circumstellar properties and molecular abundances.
We assumed a spherically symmetric envelope formed by a constant mass-loss rate driven by an accelerating wind. {Our model includes fully integrated \h2O line cooling as part of the solution of the energy balance.}}
   {We detect circumstellar molecular lines from CO, \h2O, SiO, HCN, and, for the first time in an S-type AGB star, NH\down{3}. The radiative transfer calculations result in an estimated mass-loss rate for W Aql of $4.0\e{-6} \spy$ based on the \up{12}CO lines. The estimated \up{12}CO/\up{13}CO ratio is 29, which is in line with ratios previously derived for S-type AGB stars. We find an \h2O abundance of $1.5\e{-5}$, which is intermediate to the abundances expected for M and C stars, and an ortho/para ratio for \h2O that is consistent with formation at warm temperatures. We find an HCN abundance of $3\e{-6}$, and, although no CN lines are detected using HIFI, we are able to put some constraints on the abundance, $6\e{-6}$, and distribution of CN in W Aql's circumstellar envelope { using ground-based data}. We find an SiO abundance of $3\e{-6}$, and an NH\down{3} abundance of $1.7\e{-5}$, confined to a small envelope. {If we include uncertainties in the adopted circumstellar model --- in the adopted abundance distributions, etc --- the errors in the abundances are of the order of factors of a few.} The data also suggest that, in terms of HCN, S-type and M-type AGB stars are similar, and in terms of H$_2$O, S-type AGB stars {are more like C-type than M-type} AGB stars. We detect excess blue-shifted emission in several molecular lines, possibly due to an asymmetric outflow.}
   {The estimated abundances of circumstellar HCN, SiO and \h2O place W Aql in between M- and C-type AGB stars, i.e., the abundances are consistent with an S-type classification.}


   \maketitle
%

\section{Introduction}

\begin{table*}
\caption{Molecular inventory of W Aql from HIFI observations.}             
\label{hifimol}      
\centering          
\begin{tabular}{lcrrcccc} 
\hline\hline       
Molecule	&		Transition		&	Frequency	&	$E_\mathrm{up}$	&	$\eta_\mathrm{mb}$	&	$I_\mathrm{mb}$	 &	$\theta$	& $I_\mathrm{model}/I_\mathrm{mb}$ \\
		&						& [GHz]\,\,\,\,			& [K] & & [K $\kms$]& [\arcsec]\\
\hline\hline																		CO		&	$	6 \rightarrow 5	$	&	691.473	&	116	&	0.75	&	14.1\phantom{0}	&	30.7	& 1.03\\
	&	$	10 \rightarrow 9\;\,	$	&	1151.985	&	304	&	0.64	&	14.7\phantom{0}	&	18.4	& 1.10\\
&	$	16 \rightarrow 15	$	&	1841.345	&	752	&	0.70	&	15.9\phantom{0}	&	11.5	& 1.01\\
$^{13}$CO		&	$	6 \rightarrow 5	$	&	661.067	&	111	&	0.75	&	1.3	&	32.1	& 0.87\\
&	$	10 \rightarrow 9\;\,	$	&	1101.349	&	291	&	0.74	&	\phantom{0}0.95	&	19.3	& 1.03\\
o-\h2O		&	$	1_{1,0}\rightarrow 1_{0,1}	$	&	556.936	&	61	&	0.75	&	2.0	&	38.1	& 0.68\\
	&	$	3_{1,2}\rightarrow 3_{0,3}	$	&	1097.365	&	250	&	0.74	&	2.2	&	19.3	&1.50\\
	&	$	3_{1,2}\rightarrow 2_{2,1}	$	&	1153.127	&	250	&	0.64	&	3.7	&	18.4	& 1.40\\
	&	$	3_{2,1}\rightarrow 3_{1,2}	$	&	1162.911	&	306	&	0.64	&	1.8	&	18.2	& 0.68\\
p-\h2O	&	$	1_{1,1} \rightarrow 0_{0,0}	$	&	1113.343	&	53	&	0.64	&	4.4	&	19.1	&0.87\\
HCN	&	$	13\rightarrow 12	$	&	1151.452	&	387	&	0.64	&	3.5	&	18.4	& 1.03\\
H$^{13}$CN	&	$	8 \rightarrow 7	$	&	690.551	&	149	&	0.75	&	\phantom{0}0.63	&	30.7	& 1.00\\
SiO	&	$	16 \rightarrow 15	$	&	694.294	&	284	&	0.75	&	1.1	&	30.6	& 1.07\\
$^{29}$SiO	&	$	13\rightarrow 12	$	&	557.179	&	187	&	0.75	&	\phantom{0}0.25	&	38.1	& 1.00\\
o-NH$_3$	&	$	1_0 \rightarrow 0_0	$	&	572.498	&	27	&	0.75	&	\phantom{0}0.85	&	37.1	& 1.06\\
\hline      
\end{tabular}
\tablefoot{$E_\mathrm{up}$ is the excitation energy of the upper transition level, $\eta_\mathrm{mb}$ is the main beam efficiency (see equation \ref{eta}), $I_\mathrm{mb}$ is the integrated intensity (uncertainties are taken to be 20\%), $\theta$ is the full-width half-power beam width, and $I_\mathrm{model}/I_\mathrm{mb}$ is the ratio of modelled to observed integrated intensities.}
\end{table*}

	The asymptotic giant branch (AGB) is a {late} stellar evolutionary stage of low- to intermediate-mass stars. During the AGB phase, the rate of mass loss from the surface is the main determinant of evolution and eventually leads to the final state of the star \citep{Habing1996}. The lost matter forms a circumstellar envelope (CSE) of gas from which molecules form and dust grains condense. Eventually, the matter in the CSE will be returned to the interstellar medium (ISM), thereby enriching it \citep{AGB}.
   
   There are three chemical classes of AGB stars: carbon-rich stars with C/O $> 1$ (C-type, or carbon stars), oxygen-rich stars with C/O $< 1$ (M-type), and S-type stars with C/O $\sim 1$ (hereafter called C, M, and S stars, respectively).  { The classification can be interpreted in terms of chemical evolution on the AGB, where M stars evolve into C stars through nucleosynthetic processes. Carbon is produced from He in the triple-alpha process during He-shell burning, and, through dredge-up events, is added to the stellar atmosphere. As the AGB star undergoes more dredge-ups, more carbon is added to the atmosphere and, as a consequence, M stars are transformed into C stars \citep{Herwig2005}, although this is a mass-dependent process.} In this light, the S stars are believed to be transition objects.
   
   { It has proven difficult to derive the mass-loss rate from first principles, i.e., based on a number of input stellar parameters accurately predicting the mass-loss rate of an individual star \citep{Mattsson2010}. It is therefore very important to obtain reliable estimates of the mass-loss rate from observational data.} Radiative transfer modelling of CO emission lines has been done for many AGB stars to derive their gas mass-loss rates \citep[e.g.][]{Morris1987,Kastner1992,Groenewegen1998,Schoier2001,Olofsson2002,Decin2006,Ramstedt2006,De-Beck2010}, { and it is considered to give very good estimates of this property}. { Studies of other molecular species are important for determining the physical and chemical structure (molecular as well as the solid state) of the region where the mass loss is initiated.} Molecular species that have also been studied {in sizeable samples of AGB stars} include SiO \citep{Ramstedt2009}, \h2O \citep{Maercker2009} and HCN \citep{Schoier2013}. Models describing several molecular species simultaneously for a single object have been constructed for a handful of stars such as the M star IK Tau \citep{Decin2010}, the S star $\chi$ Cyg \citep{Schoier2011}, and a small sample of C stars \citep{Woods2003}. The most well-observed and extensively modelled AGB star is the nearby C star, CW Leo (IRC+10216). As a C star, it appears to be particularly rich in circumstellar molecules with at least 60 detected to date \citep[e.g.][]{Cernicharo2000,Olofsson2008}. 
   
   W Aql is an S-type AGB star with a binary companion \citep{Ramstedt2011}. It is a Mira variable with a period of 490 days \citep{Alfonso-Garzon2012}. W Aql is one of only two S stars observed using the Heterodyne Instrument for the Far-Infrared \citep[HIFI,][]{de-Graauw2010} aboard the Herschel Space Observatory \citep{Pilbratt2010} as part of the HIFISTARS project \citep{Bujarrabal2011}. Results for the other S star, $\chi$ Cyg, are presented in \cite{Schoier2011}.
   
   It is of interest to examine how various molecular abundances vary between S stars, { and also compare them with the M and C stars}. The spectral classifications for $\chi$~Cyg and W~Aql are S6/1e and S6/6e, respectively, putting them at opposite ends of the C/O ratio scale but with similar temperatures. $\chi$~Cyg is very close to being an MS6 star and W~Aql is one C/O grade away from being an SC star \citep{Keenan1980}. As such, they present an interesting demonstration of how S stars may vary as they transition from being more oxygen-rich to more carbon-rich.

   { In this paper we present new spectral line observations of W Aql made by the HIFI and PACS instruments (\sec{obs:hifi} and \sec{obs:pacs}, respectively) on the Herschel Space Observatory, which we complement with ground-based observations (\sec{obs:ground})}. We perform a detailed excitation analysis for all the observed molecules to determine the circumstellar characteristics of the star (\sec{sec:mod}), including mass-loss rate, temperature structure, and molecular abundances. We also compare our results with those of other AGB stars (\sec{sec:dis}).

\section{Observations}\label{sec:obs}

\subsection{HIFI Observations}\label{obs:hifi}

   \begin{figure*}
   \centering
      \includegraphics[width=\textwidth]{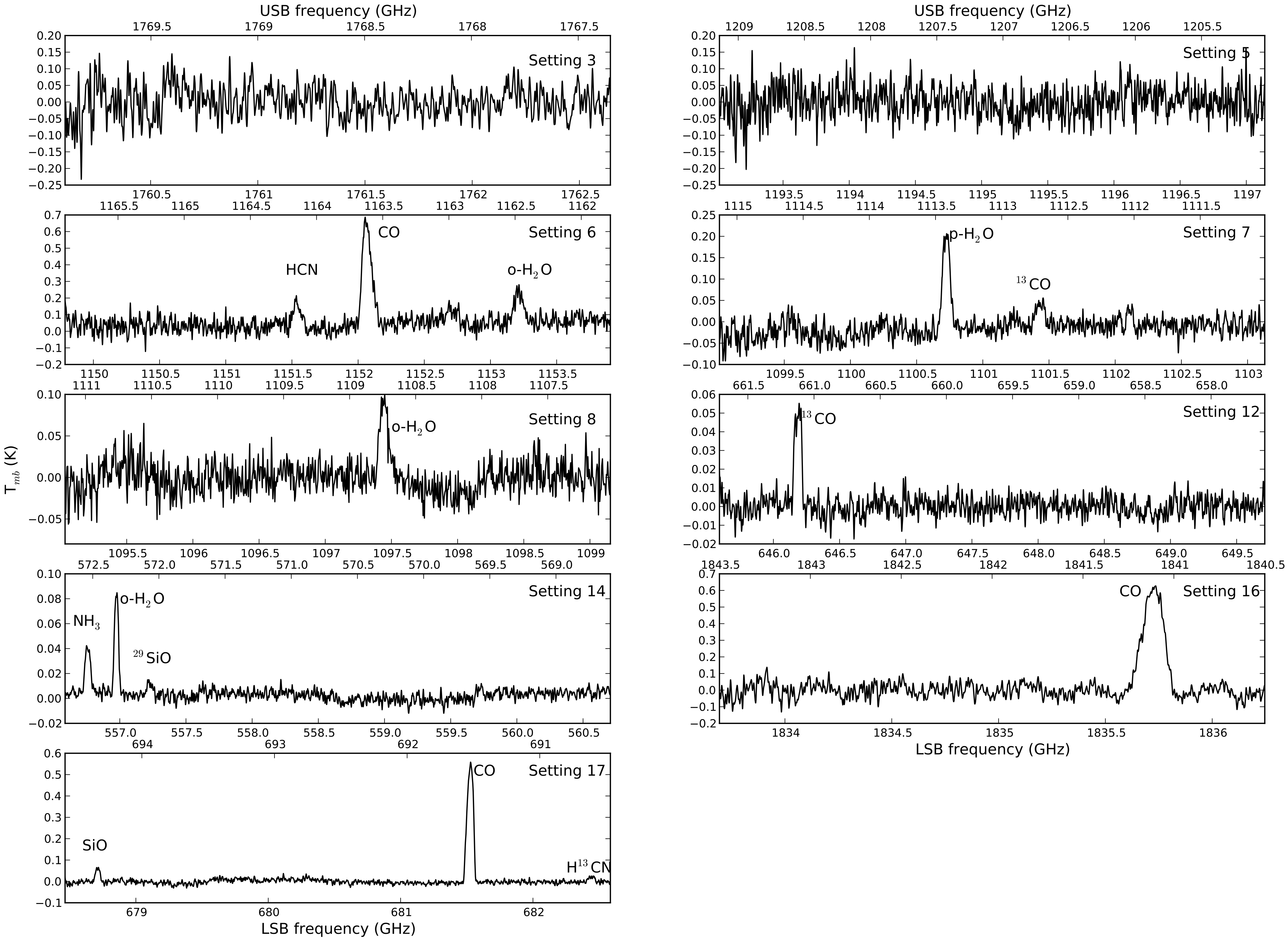}
   \caption{The spectra of W Aql as observed using HIFI. The top right corner of each spectrum gives the frequency setting within the HIFISTARS project. Strong lines are labelled with their corresponding molecule identification. {The spectra are obtained in double sideband mode, and the lower sideband (LSB) and upper sideband (USB) frequencies are given below and above the spectra, respectively.} The spectra are corrected for baseline effects (see text). The intensity scale is given in main beam brightness temperature.}
              \label{HIFI}%
    \end{figure*}

\begin{table}[t]
\caption{HIFISTARS settings}
\label{hifisets}
\begin{center}
\begin{tabular}{cccc}
\hline\hline
Setting & LO & LSB range & USB range\\
 & [GHz] & [GHz] & [GHz]\\
\hline
\phantom{*}3* & 1764.997 & 1760.09 -- 1762.66 & 1767.34 -- 1769.91\\
5 & 1201.081 & 1193.01 -- 1197.17 & 1204.99 -- 1209.14\\
6 & 1157.847 & 1149.79 -- 1153.92 & 1161.77 -- 1165.90\\
7 & 1107.071 & 1098.99 -- 1103.14 & 1111.00 -- 1115.15\\
8 & 1103.093 & 1095.04 -- 1099.17 & 1107.02 -- 1111.15\\
12 & \phantom{0}653.625 & 645.58 -- 649.75 & 657.58 -- 661.71\\
14 & \phantom{0}564.648 & 556.59 -- 560.72 & 568.58 -- 572.71\\
16 & 1838.598 & 1833.68 -- 1836.26 & 1840.94 -- 1843.52\\
17 & \phantom{0}686.526 & 678.46 -- 682.61 & 690.44 -- 694.59\\
\hline
\end{tabular}
\end{center}
\tablefoot{LO is the local oscillator frequency, LSB and USB are the lower and upper sidebands, respectively. * indicates that the frequency range observed for W Aql was not the same range that was proposed or observed as setting 3 for other HIFISTARS sources.}
\end{table}%
    
\begin{table}[tb]
\caption{ PACS line observations of CO, \h2O, HCN, and NH$_3$.}
\label{pacslines}
\centering         
\begin{tabular}{crrcc} 
\hline\hline       
		Transition	& Peak $\lambda$& $E_\mathrm{up}$ &		$F_{\mathrm{obs}}$ & $\frac{F_{\mathrm{mod}}}{F_{\mathrm{obs}}}$\\
							& $[\mic]\;$&[K]			&[W m$^{-2}$]\\
\hline
CO\\
 $14\rightarrow13$ & 185.991&580 & $(3.7\pm0.7)\times 10^{-16}$ & 0.89\\
 $15\rightarrow14$ & 173.629&663 & $(3.5\pm0.7)\times 10^{-16}$ & 0.98\\
 $16\rightarrow15$ & 162.810&752 & $(4.2\pm0.8)\times 10^{-16}$ & 0.85\\
 $17\rightarrow16$ & 153.271&847 & $(4.8\pm1.0)\times 10^{-16}$ & 0.80\\
 $18\rightarrow17$ & 144.785&946 & $(4.9\pm1.0)\times 10^{-16}$ & 0.80\\
 $19\rightarrow18$ & 137.218&1051 & $(4.5\pm0.9)\times 10^{-16}$ & 0.89\\
 $20\rightarrow19$ & 130.387&1161 & $(5.2\pm1.0)\times 10^{-16}$* & 0.80\\
 $21\rightarrow20$ & 124.196&1277 & $(5.0\pm1.0)\times 10^{-16}$ & 0.85\\
 $22\rightarrow21$ & 118.586&1398 & $(5.1\pm1.0)\times 10^{-16}$ & 0.85\\
 $24\rightarrow23$ & 108.761&1657 & $(4.2\pm0.8)\times 10^{-16}$ & 1.06\\
 $25\rightarrow24$ & 104.451&1796 & $(4.1\pm0.8)\times 10^{-16}$ & 1.10\\
 $28\rightarrow27$ &93.349 &2242 & $(3.0\pm0.6)\times 10^{-16}$* & ...\\
 $29\rightarrow28$ & 90.156&2402 & $(3.3\pm0.7)\times 10^{-16}$ & ...\\
 $31\rightarrow30$ & 84.410& 2737 & $(2.9\pm0.6)\times 10^{-16}$ & ...\\
 $32\rightarrow31$ & 81.813&2913 & $(2.0\pm0.4)\times 10^{-16}$ & ...\\
 $33\rightarrow32$ & 79.363 &3095 & $(1.7\pm0.4)\times 10^{-16}$ & ...\\
 $34\rightarrow33$ & 77.062&3282 & $(1.8\pm0.4)\times 10^{-16}$ & ...\\
 $36\rightarrow35$ & 72.856&3672 & $(1.1\pm0.2)\times 10^{-16}$ & ...\\
o-\h2O\\
$2_{2,1}\rightarrow 2_{1,2}$ & 180.479&194 &$(5.7\pm0.1) \times 10^{-16}$ & 1.24\\
 $2_{1,2}\rightarrow 1_{0,1}$& 179.500&114 & $(2.0\pm0.4)\times 10^{-16}$ & 0.87\\
$5_{1,4}\rightarrow 5_{0,5}$& 134.423&575 & $(9.5\pm0.2)\times 10^{-17}$ & 0.78\\
 $4_{2,3}\rightarrow 4_{1,4}$&132.423& 432 & $(1.1\pm0.2)\times 10^{-16}$ & 0.81\\
 $4_{3,2}\rightarrow 4_{2,3}$& 121.721&552 & $(9.2\pm2.0)\times 10^{-17}$ & 0.76\\
 $2_{2,1}\rightarrow 1_{1,0}$& 108.096&194 & $(4.1\pm0.8)\times 10^{-16}$ & 0.74\\
 $6_{1,6}\rightarrow 5_{0,5}$& 82.030&644 & $(3.1\pm0.7)\times 10^{-16}$ & 1.12\\
 $4_{2,3}\rightarrow 3_{1,2}$& 78.747 &552 & $(4.7\pm0.9)\times 10^{-16}$ & 0.93\\
 $3_{2,1}\rightarrow 2_{1,2}$& 75.389 &305 & $(5.4\pm1.1)\times 10^{-16}$ & 0.98\\
p-\h2O\\
 $3_{1,3}\rightarrow 2_{0,2}$& 138.530&285 & $(2.2\pm0.4)\times 10^{-16}$* & 0.98\\
 $4_{0,4}\rightarrow 3_{1,3}$& 125.376&320 & $(1.8\pm0.4)\times 10^{-16}$ & 1.09\\
 $5_{2,4}\rightarrow 5_{1,5}$& 111.626&599 & $(6.9\pm0.2)\times 10^{-16}$ & 1.11\\
 $3_{2,2}\rightarrow 2_{1,1}$& 89.983 &297 & $(2.9\pm0.6)\times 10^{-16}$ & 1.07\\
 $3_{3,1}\rightarrow 2_{2,0}$& 67.095&410 & $(3.6\pm0.7)\times 10^{-16}$ & 1.23\\
HCN \\
 $18\rightarrow 17$ &188.145& 728 & $(6.0\pm1.5)\e{-17}$ & 0.84\\
 $19\rightarrow 18$ &178.264& 808 & $(6.7\pm1.6)\e{-17}$ & 0.76\\
 $20\rightarrow 19$ &169.390& 893 & $(4.9\pm1.2)\e{-17}$ & 1.06\\
 $21\rightarrow 20$ &161.368& 983 & $(9.7\pm2.1)\e{-17}$ & 0.53\\
 $22\rightarrow 21$ & 154.042&1076 & $(7.1\pm1.8)\e{-17}$ & 0.72\\
 $23\rightarrow 22$ & 147.363&1174 & $(8.4\pm1.8)\e{-17}$ & 0.59\\
 $24\rightarrow 23$ & 141.277&1276 & $(6.7\pm1.5)\e{-17}$ & 0.70\\
 $25\rightarrow 24$ & 135.636&1382 & $(7.2\pm1.7)\e{-17}$ & 0.59\\
 $27\rightarrow 26$ & 125.647&1607 & $(5.9\pm1.7)\e{-17}$ & 0.60\\
 $28\rightarrow 27$ & 121.186&1726 & $(5.3\pm1.2)\e{-17}$ & 0.66\\
p-NH\down{3}\\
 $3_2\rightarrow2_2$ & 169.979& 150 & $(1.8\pm0.4)\e{-16}$ & 0.73\\
 $4_2\rightarrow3_2$ & 127.168& 264 & $(2.4\pm0.5)\e{-16}$&1.14\\
 $7_5\rightarrow6_5$ & 72.530&666 & $(6.5\pm2.1)\e{-16}$&0.58\\
 $9_4\rightarrow8_4$ & 56.340&1196 & $(3.1\pm0.7)\e{-16}$& 1.12\\
\hline
\end{tabular}
\tablefoot{Peak $\lambda$ is the observed central wavelength of the detected line, $F_\mathrm{obs}$ is the observed line flux, and $\frac{F_{\mathrm{mod}}}{F_{\mathrm{obs}}}$ is the ratio of modelled to observed flux. An * in the $F_\mathrm{obs}$ column indicates that the line is known to be blended with a relatively faint \h2O or HCN line. These remain in our analysis as they should not significantly alter results.}
\end{table}

W Aql was observed in September and October of 2011 using the HIFI \citep{de-Graauw2010} instrument\footnote{https://herschel.jpl.nasa.gov/hifiInstrument.shtml} aboard the Herschel Space Observatory as part of the HIFISTARS Guaranteed Time Key Programme \citep{Bujarrabal2011}. It was observed in nine different frequency settings. The spectra are presented in Fig. \ref{HIFI}, labelled according to the frequency setting within the HIFISTARS project, and with the carrier of the most prominent lines indicated. The local oscillator frequencies and frequency ranges for each setting are listed in Table \ref{hifisets}. 

The data were processed using HIPE\footnote{http://www.cosmos.esa.int/web/herschel/data-processing-overview} \citep[version 8,][and note that the HIFI calibration files have remained stable with no significant changes]{Ott2010}, the standard Herschel data pipeline. Each band was observed in two orthogonal linear polarisations (vertical and horizontal) and these were averaged in HIPE. The exception was setting 3, for which a ripple effect \citep[thought to be a product of standing waves in the hot electron bolometer mixers, see][]{Roelfsema2012} was present in the vertical polarisation. In this case, only the horizontal polarisation was used. Polynomials of order up to two were used for baseline subtraction. { Some baseline effects remain, but they do not affect our line measurements significantly.}

A variety of molecules, CO, \h2O, HCN, SiO, and NH$_3$, were identified (see Fig \ref{HIFI}). All the detected molecular lines are listed in Table \ref{hifimol}. No lines were detected in settings 3 and 5. We believe this is the first detection of NH\down{3} in an S star, although NH\down{3} has been previously detected in C and M stars, see for example \citet{Hasegawa2006} and \citet{Menten2010}.

After processing using HIPE, and extracting profiles for individual emission lines, the line intensities were corrected for the main beam efficiency \citep{Roelfsema2012},

\begin{equation}\label{eta}
\eta_\mathrm{mb} = \eta_\mathrm{mb,0} \exp \left[ -(4\pi\sigma/\lambda)^2\right],
\end{equation}
where $\sigma = 3.8 \um$ is the telescope surface accuracy, $\lambda$ the wavelength of the transition (in $\um$), and $\eta_\mathrm{mb,0}$ is 0.66 in HIFI band 5 (1120 -- 1280 GHz) and 0.76 for all other frequencies. The correction factors $\eta_\mathrm{mb}$ are given in Table \ref{hifimol}, as are the full-widths at half-power, $\theta$, for HIFI.

{We chose the final velocity resolution of the individual lines depending on their signal-to-noise ratios. Normally this is to a velocity resolution better than $2\;\kms$, except for some of the weak lines from the rarer isotopologues.}

\subsection{PACS Observations}\label{obs:pacs}

{
W Aql was observed with PACS\footnote{https://herschel.jpl.nasa.gov/pacsInstrument.shtml} \citep{Poglitsch2010} in SED mode on November 6 and 7, 2010, in observation identifiers (obsids) 1342209731 (band B2A: 55--70 $\mic$, band R1A: 102--146 $\mu$m) and 1342209732 (band B2B: 70-95 $\mic$, band R1B: 140-210 $\mu$m), in chopping/nodding mode, with a chopper throw of 3 arcminutes. Each wavelength range scan was performed twice. Including the calibration blocks ($\sim$\,two minutes / obsid), the duration of the two obsids were 2124 and 4065 seconds, respectively.

The data reduction was performed in HIPE 12, with the standard data reduction pipeline for SED-range spectroscopy. The data were rebinned with an oversampling factor of 2, which corresponds to Nyquist sampling with respect to the instrumental resolution. The flux calibration (absolute and relative) was performed via normalisation to the telescope background. PACS calibration set 65 was used throughout the process.   

After level 2 (one spectrum / spatial pixel), W Aql was considered a point source, so we applied a beam correction to the central spaxel. We then compared the obtained continuum with the one resulting from the integral of all spaxels, and found both consistent over the whole wavelength range.

We searched the calibrated PACS spectrum for lines corresponding to molecules identified in the HIFI scans. We found many lines corresponding to CO, H$_2$O, HCN and NH$_3$, which are listed in Table \ref{pacslines}. The line strengths were measured by fitting a Gaussian on top of a continuum \citep{Lombaert2013}. The reported uncertainties include the fitting uncertainty and the absolute-flux-calibration uncertainty of 20\%. We excluded any lines which were obvious line-blends according to one of two criteria: 1) the FWHM of the fitted Gaussian was larger than the FWHM of the PACS spectral resolution by at least 20\%, 2) multiple CO or H$_2$O transitions had a central wavelength within the FWHM of the fitted central wavelength of the emission line. Such blends were excluded from our analysis. We caution the reader that the line strengths in Table 2 may still be affected by emission from other molecules or transitions not included in our line list. Our method still leaves approximately half of the detected PACS lines unidentified. The unidentified and blended lines are listed in Table \ref{Ulines}. We did not find any convincing detections of SiO or $^{13}$CO lines.

}
\subsection{Complementary observations}\label{obs:ground}

We used previously published line observations of CO, SiO, HCN, and CN from \citet{Bachiller1997}, \citet{Ramstedt2009}, \cite{De-Beck2010} and \citet{Schoier2013} as listed in Table \ref{otherlines} to constrain our models further.

Our spectral energy distribution (SED) model was fitted to photometry from IRAS, 2MASS \citep{Hog2000}, and spectra from the Infrared Space Observatory \citep[ISO,][]{Kessler1996} Short and Long Wavelength Spectrometers (SWS and LWS, respectively), processed by \cite{Sloan2003} for the SWS spectrum and extracted from the archive for the LWS spectrum.

\begin{table*}[tb]
\caption{Complementary line observations.}
\label{otherlines}
\centering          
\begin{tabular}{lcrcrrcc} 
\hline\hline       
	&		Transition	& $E_\mathrm{up}$ &	Telescope	&	$I_\mathrm{mb}\;\;\;\;\;$	&	$\theta\;$	 & Ref & $I_\mathrm{model}/I_\mathrm{mb}$\\
	&					& [K]			& &[K $\kms$] & [\arcsec]\\
\hline
CO 	& $1\rightarrow0$ & 6 & SEST & 29.6 & 45 & D10 & 1.01\\
	& $2\rightarrow1$ & 17 & APEX & 64.9 & 27 & A & 1.12\\
	& $2\rightarrow1$ & 17 & JCMT & 79.2 & 20 & D10 & 1.51\\
	& $3\rightarrow2$ & 33 & APEX & 136\phantom{.0} & 18 & R09 & 0.80\\
	& $4\rightarrow3$ & 55 & APEX & 130\phantom{.0} & 14 & A & 0.97\\
	& $4\rightarrow3$ & 55 & JCMT & 200\phantom{.0} & 11 & R09 & 1.01\\
	& $7\rightarrow6$ & 155 & APEX & 162\phantom{.0} & 8 & A & 1.03\\	
$^{13}$CO	& $2\rightarrow1$ & 16 & APEX & 6.4 & 28 & A & 0.47\\
	& $3\rightarrow2$ & 32 & APEX & 10\phantom{.0} & 18 & A & 0.68\\
SiO & $2\rightarrow1$ & 6 & OSO & 2.6 & 44 & R09 & 1.11\\
	& $3\rightarrow2$ & 13 & SEST & 3.9 & 45 & R09 & 0.89\\
	& $5\rightarrow4$ & 31 & APEX & 6.4 & 28 & A & 1.18\\
	& $8\rightarrow7$ & 75 & APEX & 10.3 & 14 & A& 1.92\\
	& $11\rightarrow10$ & 138 & APEX & 15.1 & 13 & A& 0.84\\
HCN & $1\rightarrow0$ & 4 & IRAM & 18.0 & 29 & S13 & 1.02\\
	& $3\rightarrow2$ & 26 & IRAM & 63.4 & 9 & S13 & 1.95\\
	& $4\rightarrow3$ & 43& APEX & 25.5 & 18 & S13 & 1.03\\
CN	& $1_{1/2}\rightarrow0_{1/2}$ & 5 & IRAM & 8.1 & 22 &B97 & 0.66\\
	& $1_{3/2}\rightarrow0_{1/2}$ & 5 & IRAM & 8.9 & 22 &B97 & 1.07\\
	& $2_{3/2}\rightarrow1_{1/2}$ & 16 & IRAM & 11.0 & 12 &B97& 0.93\\
	& $2_{5/2}\rightarrow1_{3/2}$ & 16 & IRAM & 13.5 & 12 &B97 & 1.35\\
\hline
\end{tabular}
\tablefoot{References: A = data from the ESO archive, R09 = \citet{Ramstedt2009}, D10 = \citet{De-Beck2010}, S13 = \citet{Schoier2013}, B97 = \cite{Bachiller1997}.}
\end{table*}

\section{W Aql}

For the purposes of our radiative transfer modelling, we have used a distance to W Aql of 395 pc, derived from \citeapos{Whitelock2008} period-magnitude relation and the 2MASS K band magnitude \citep{Cutri2003}. At this distance, the angular separation of W Aql and its companion, which is $0\farcs46$ \citep{Ramstedt2011}, corresponds to a projected distance of 190 AU. {This could mean that the companion is moving through the inner part of the CSE, but it is also possible that the companion lies behind or in front of the CSE. We have performed an optical spectroscopic classification of the companion \citep[discussed in detail in][]{Danilovich2014a} and found it to be an F8V star, mostly likely lying behind the CSE. The lack of significant disruption to the molecular emission lines strongly suggests that the companion is not as close as 190 AU to the AGB star in real terms. Furthermore, \cite{Mayer2013} used photometric methods to classify the companion and find results consistent with a K4 star, which additionally indicates that the companion is suffering from extinction effects due to the dust produced by the AGB star,} and hence is most likely within or behind the CSE. {We calculated the luminosity of W Aql using \citeapos{Glass1981} period-luminosity relation, which gives a luminosity of 7500 $\lsol$.}

\section{Radiative transfer modelling}\label{sec:mod}

{
We use four different codes in our modelling process. Each of these is discussed in detail in the rest of this section. An outline of our modelling process follows.

We begin by estimating some of the dust properties using an SED model, which gives us the stellar temperature and the dust mass-loss rate. This is fed in to the dynamical model, which gives us an estimate of the dust and gas velocity profiles and the total mass-loss rate. Together they give us an estimate of the dust-to-gas mass ratio.

The results of these models are used as input in our CO line modelling using a radiative transfer code based on the Monte Carlo method (MCP). Once we have a reasonable CO model which fits the observed line intensities (with a converged temperature profile), we use the same parameters in our radiative transfer code based on the accelerated lambda iteration method (ALI) to run an \h2O line model. The \h2O modelling outputs an \h2O line cooling profile, which we feed back into the CO model. We then adjust the CO model to account for the additional cooling, and run the \h2O modelling again using the updated results on mass-loss rate and gas temperature structure. After a few iterations between the CO line model in MCP and the \h2O line model in ALI, we converge on a consistent description of the CSE gas temperature structure and other parameters. These parameters are then used for the radiative transfer modelling of the other molecules using MCP.
}

\subsection{SED and dynamical modelling}\label{sec:seddyn}

The radiative transfer through the dust shell is based on \cite{Haisch1979}, which allows for a size distribution of the dust grains. The SED model we employed assumes that the main dust component is due to amorphous silicate dust based on the properties described in \cite{Justtanont1992}. It has as input the distance to the star, the gas terminal expansion velocity (measured from CO line data), the dust grain density (taken to be 3.3 g cm\up{-3}), and the dust condensation temperature (taken to be 1000 K for silicate dust). {We assume spherical grains with an $a^{-3.5}$ size distribution following \cite{Mathis1977}, with radii ranging from $5\e{-3}$ to $0.25 \um$. We do not attempt to fit the specific dust features but only the overall SED to get an estimate of the dust radiation field (essential for the excitation of the molecular lines), hence we chose silicate dust as a proxy for the dust around W Aql although as an S star the dust composition is likely to be more varied. We are not attempting to determine the composition of the dust around W Aql.}

\begin{figure}[tb]
   \centering
   \includegraphics[width=8.8cm]{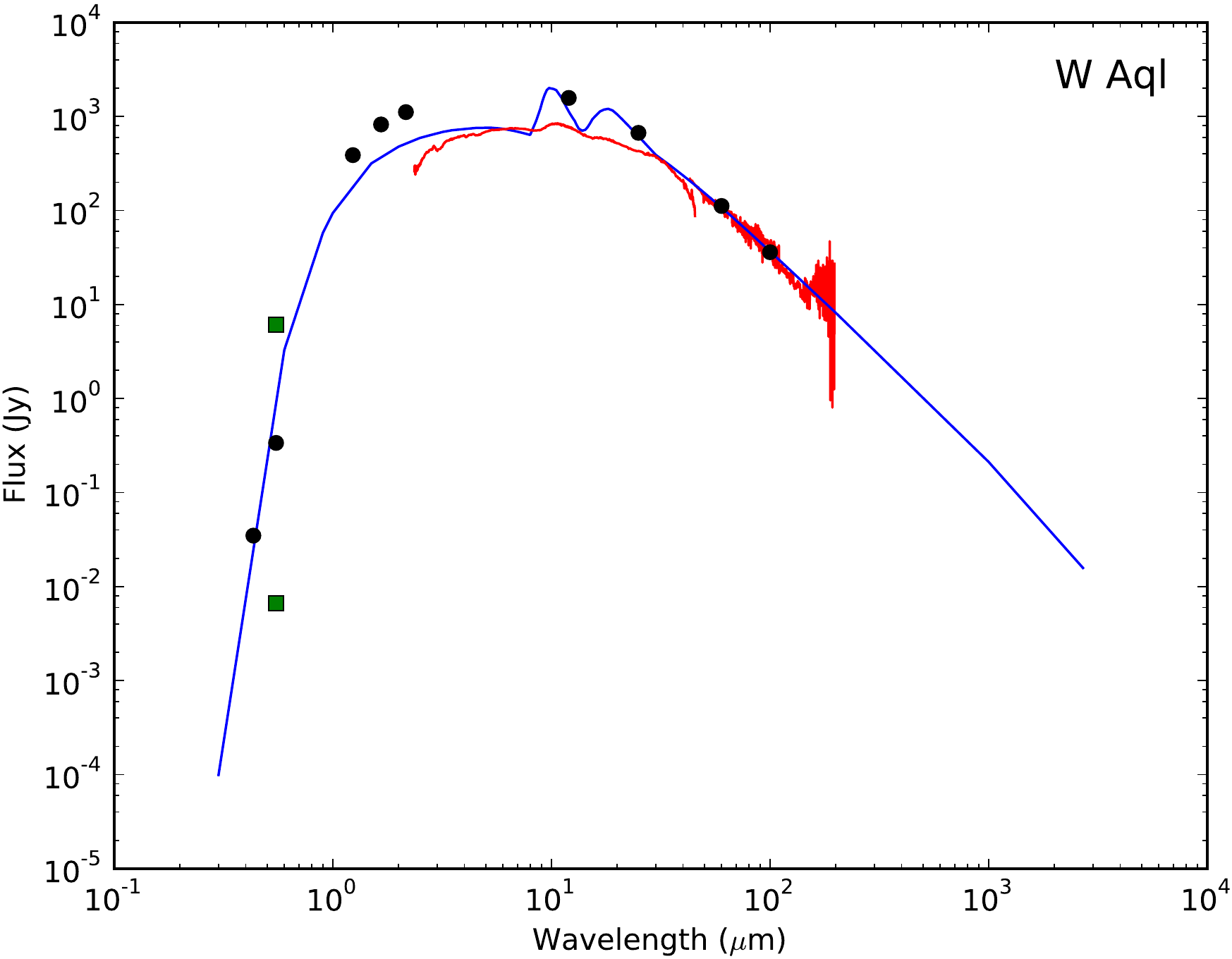}
      \caption{The SED model (blue line) shown with the ISO spectral scans (red) and photometric observations (black points). The green squares show W Aql's variability in the V band.}
         \label{dustSED}
   \end{figure}

\begin{figure}[tb]
   \centering
   \includegraphics[width=8.8cm]{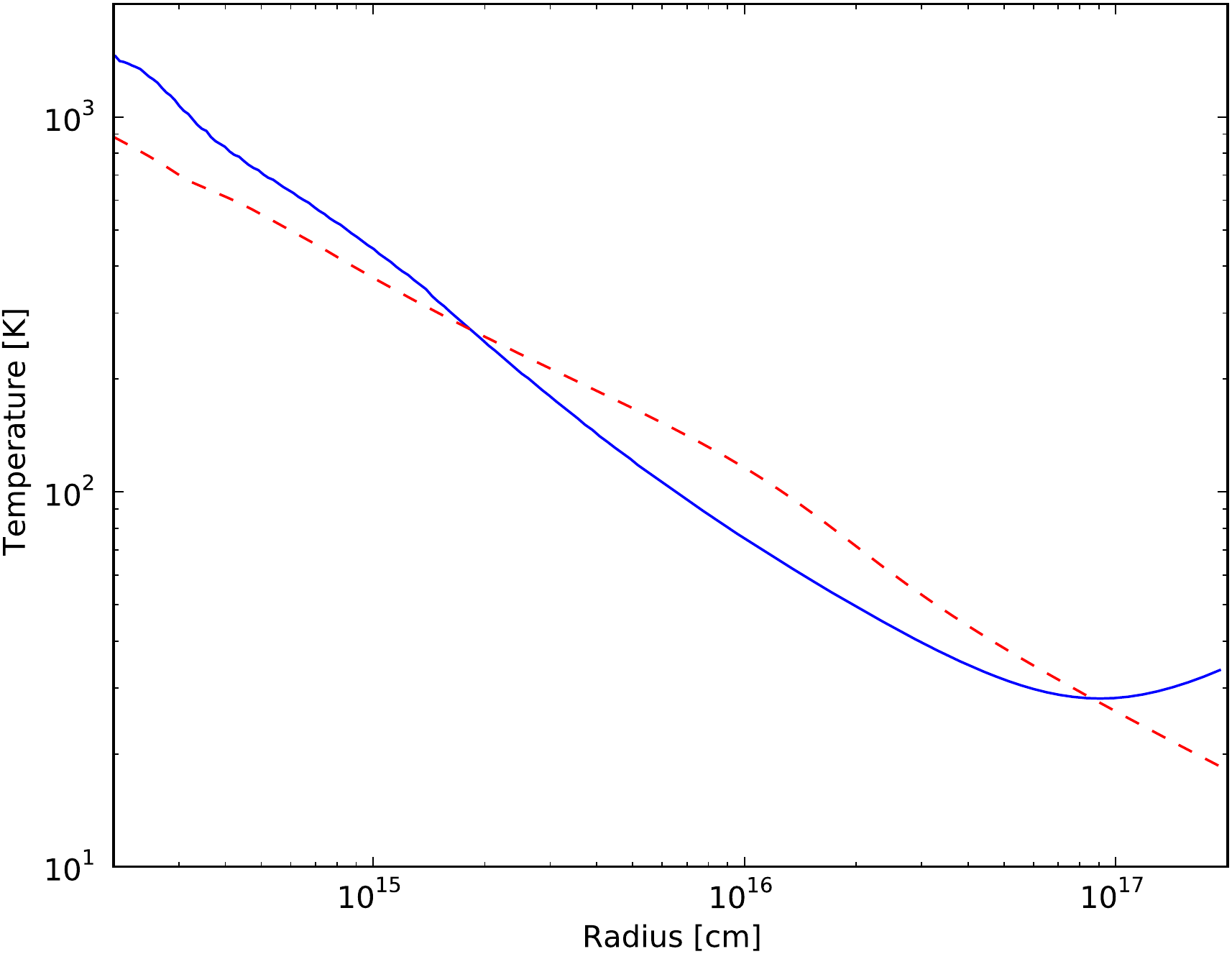}
      \caption{The dust temperature profile obtained from the SED modelling (red dashed line) and the gas temperature profile obtained from the CO line modelling (blue solid line).}
         \label{temp}
   \end{figure}

{ The SED model was fit to the photometric and ISO observations and the stellar temperature and the dust optical depth were estimated from the model}. As W Aql is a Mira variable, there is variation across the pulsation cycle, for example in temperature and luminosity. This leads to significant variation between measurements taken at different times. The variation is visible in Fig. \ref{dustSED}, in the difference between the 2MASS K photometric point and the $2\um$ end of the ISO SWS spectrum. The effect increases with decreasing wavelength as indicated by the two green squares taken from the minimum and maximum  V magnitudes in the AAVSO database, where there is a variation of three orders of magnitude in flux. The resulting model SED can be seen in Fig. \ref{dustSED} and the results are summarised in Table \ref{dynSED}. The dust temperature profile is shown in Fig. \ref{temp}. { The time variability of the stellar flux could potentially affect the excitation of the molecules. We are presently unable to take into account that the observational line data are taken at different epochs (and therefore at different stellar flux levels) and, in this sense, we produce an average model.}

After finding the dust mass-loss rate from the SED fitting, a dynamical model was applied, assuming a dust-driven wind and {the same type of dust}. The dynamical modelling is based on solving the dust and gas velocity equations simultaneously, with the observed gas expansion velocity and the results from the SED modelling as constraints, as described in \cite{Ramstedt2008}. The mass of W Aql is assumed to be 1\,$\msol$. 
The dust-to-gas mass ratio, $\Psi$, was tuned until the {terminal} gas velocity of the dust-driven wind model matched the observations.
The results are an estimate of the total mass-loss rate, the gas and dust velocity radial profiles, and, as a consequence, the drift velocity radial profile. We found $\Psi = 2\e{-3}$. { This value was then used as a starting point in our line modelling, where $\Psi$ enters into the gas-dust heating term as discussed in \sec{COmod}}. {The model's sensitivity to the initial mass is moderate; a 50\% change in mass corresponds to a 10\% change in the drift velocity --- defined as the difference between the dust and the gas velocity --- which is an important parameter in the energy balance.}

However, in the line fitting, it became evident that the gas velocity radial profile obtained from the dynamical model did not lead to a good fit to the observed CO lines (the gas acceleration is too high in the inner CSE). Hence, a gas velocity radial profile is adopted with one free parameter, which is adjusted to fit the line shapes:
\begin{equation}\label{veleq}
\upsilon(r) = \upsilon_\mathrm{min} + \left(\upsilon_\infty - \upsilon_\mathrm{min}\right)\left( 1 - \frac{R_\mathrm{in}}{r}\right)^\beta
\end{equation}
where $\upsilon_\infty$ is the final expansion velocity, $R_\mathrm{in} = 2\e{14}$ cm is the dust condensation radius (i.e., the innermost radius for which acceleration takes place), and $\upsilon_\mathrm{min}$ the minimum velocity taken as the sound speed ($3\;\kms$) at $R_\mathrm{in}$. 

A profile of this form fits the dynamical model result with $\beta = 0.7$ { compared with $\beta = 2.0$ from fitting the lines. We further adopt a constant drift velocity of 9.5 $\kms$ throughout the CSE, which is the terminal value given by the dynamical model. The velocity profiles that were subsequently used in our line modelling are shown in Fig. \ref{vels}.}

\begin{table}
\caption{Parameters of W Aql and its CSE obtained from the SED and dynamical modelling, and through a cursory inspection of the line data.}
\label{dynSED}      
\centering                          
\begin{tabular}{l r}        
\hline\hline                 
Distance		& 	395 pc\\
Effective temperature, $T_{\mathrm{eff}}$ & 2300 K\\
Dust/gas mass ratio, $\Psi$ & $2 \e{-3}$\\
Inner radius of dust shell & $2\e{14}$ cm\\
Dust optical depth at $10\um$ & 0.6\\
n(CO)/n(\h2) (adopted) & $6\e{-4}$\\
Luminosity, $L_*$ & 7500 $\lsol$\\
Stellar mass (adopted) $M_*$ & 1 $\msol$\\
Stellar velocity, $\upsilon_\mathrm{LSR}$ & $23 \;\kms$\\
Gas expansion velocity, $\upsilon_\infty$ & $16.5\;\kms$\\
\hline                                   
\end{tabular}
\end{table}


\subsection{The circumstellar model}

The CSE is assumed to have spherical symmetry due to a constant, isotropic, and smooth mass-loss rate. { This is a first order approximation since it is known that at least the dust CSE of W Aql shows non-spherical features \citep{Ramstedt2011}.} The dominant gas component is assumed to be molecular hydrogen.

\begin{figure}[tb]
   \centering
   \includegraphics[width=8.8cm]{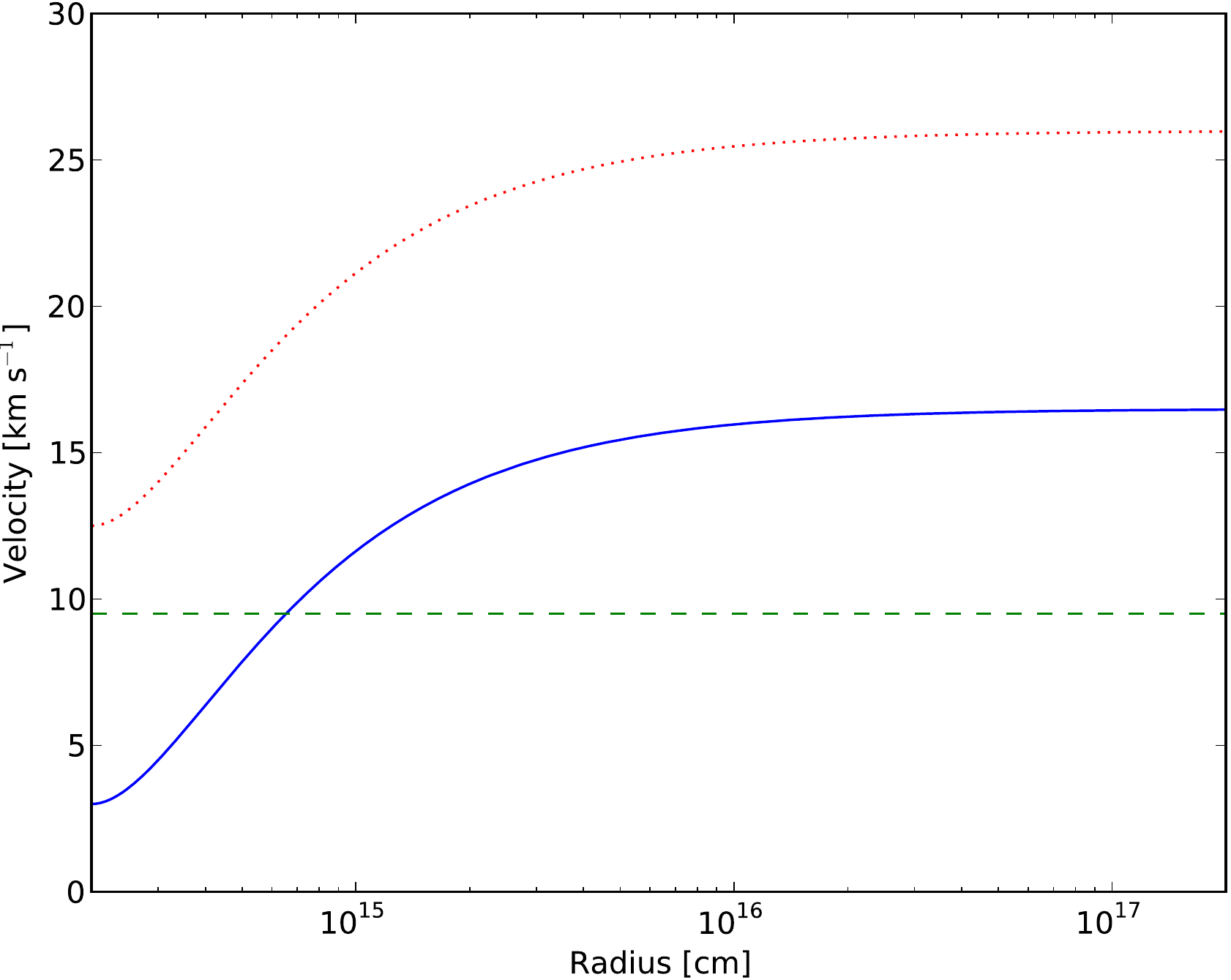}
      \caption{The gas (blue solid line), dust (red dotted line) and drift (green dashed line) velocity radial profiles as determined from the CO line fitting.}
         \label{vels}
   \end{figure}

We also assume a smoothly accelerating wind as there is a clear difference in widths between the observed low- and high-$J$ CO lines in W Aql. The radial gas velocity law we employed to account for this is given in \eq{veleq}. As will be shown, a value of $\beta = 2$ gives a good fit to the line shapes {(based on resolved CO and \h2O lines)}, where the higher-energy lines are noticeably narrower than the lower-energy lines. The gas velocity profile resulting from this fit can be seen in Fig. \ref{vels} together with the dust and drift velocities. 

\subsection{The best-fit criterion}

The best fit models were found using a $\chi^2$ statistic defined as
\begin{equation}
\chi^2 = \sum_{1}^N \frac{(I_\mathrm{mod} - I_\mathrm{obs})^2}{\sigma^2}
\end{equation}
where $I$ is the integrated line intensity, $N$ the number of observed lines, and $\sigma$ the uncertainty on the intensity measurements, generally assumed to be 20\% unless otherwise noted. The reduced $\chi^2$ value is given by $\chi^2_\mathrm{red} = \chi^2/(N-p)$ where $p$ is the number of free parameters, generally $p=1$, except for the CO and NH\down{3} models where $p=2$. Our cited errors are for a 90\% confidence interval where multiple observed lines are available. For models of a single observed line, the uncertainty we give is based on a 20\% uncertainty in the observed line intensity.

\subsection{CO line modelling}\label{COmod}

   \begin{figure*}[tb]
   \centering
   \includegraphics[width=8.8cm]{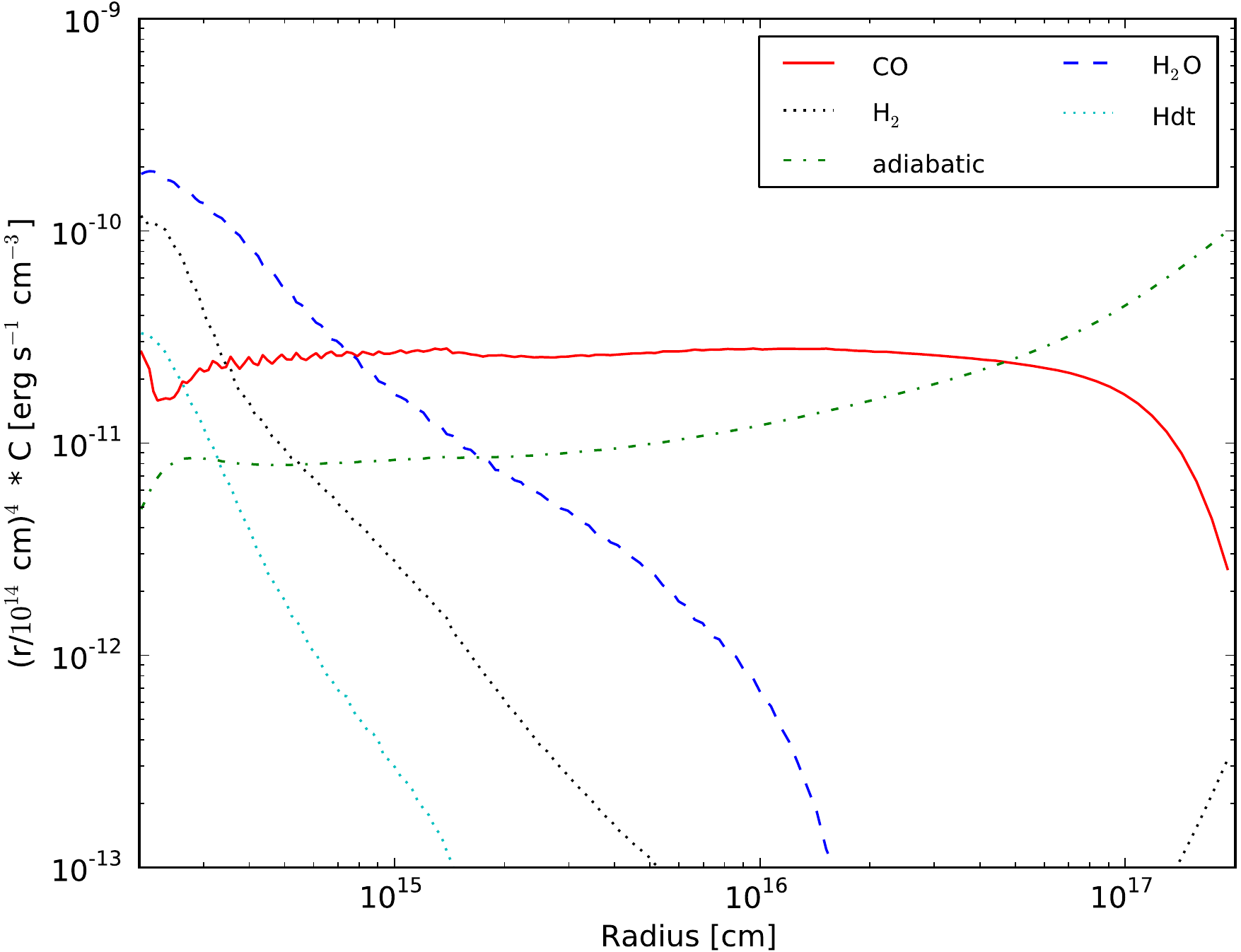}
   \includegraphics[width=8.8cm]{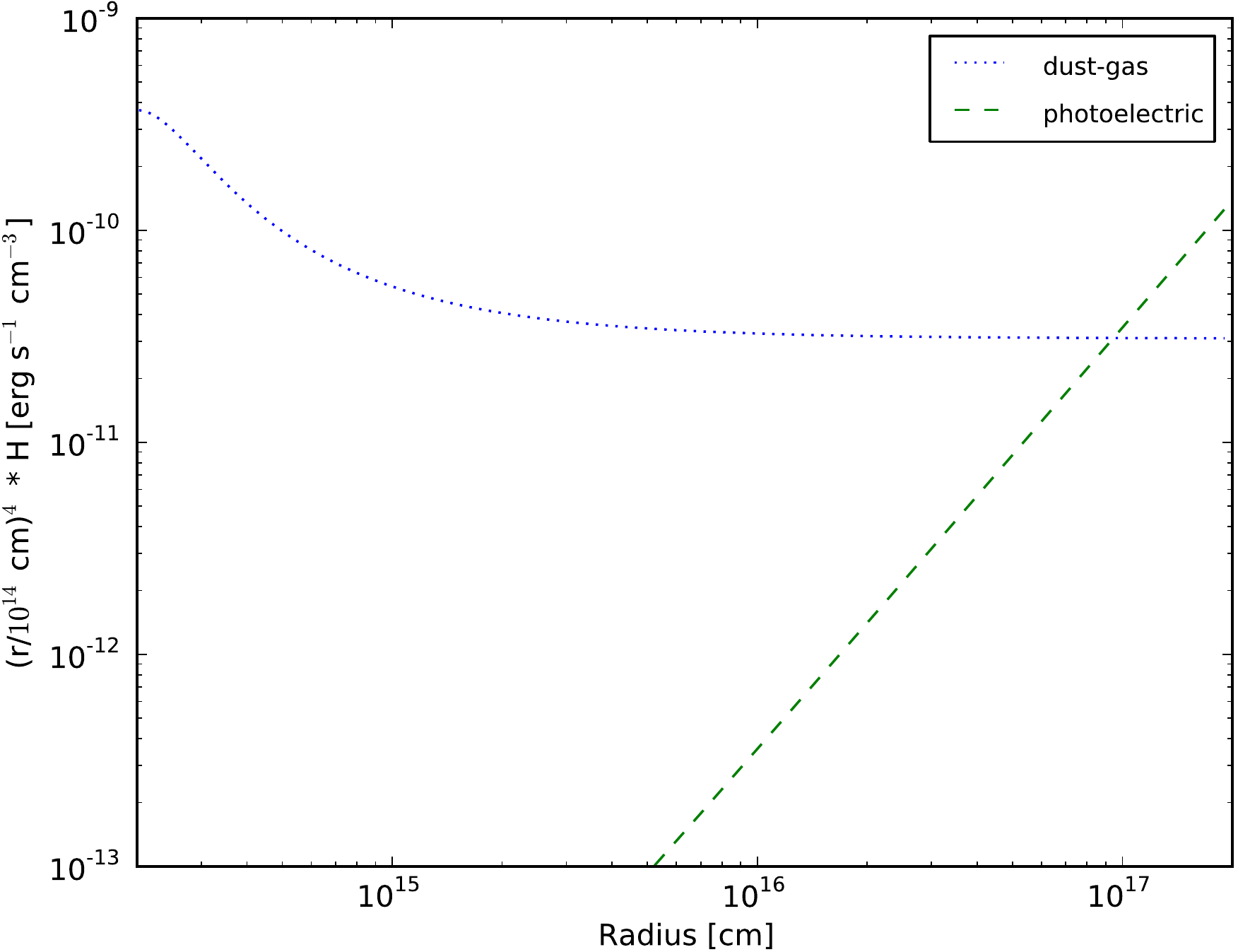}
      \caption{The gas cooling (\textit{left}) and heating (\textit{right}) rates determined in the best-fit CO model. The cooling includes contributions from CO, \h2O and \h2 line cooling, adiabatic cooling, and cooling due to non-collisional interactions of gas and dust (Hdt). The heating includes some CO line heating (when the cooling is negative), heating from dust-gas collisions, and photoelectric heating from the interstellar radiation field. {There is a slight ``wiggle" in the CO cooling function due to unavoidable Monte-Carlo noise. This does not have an impact on our results.}}
         \label{heatcool}
   \end{figure*}

   \begin{figure}[tb]
   \centering
   \includegraphics[width=8.8cm]{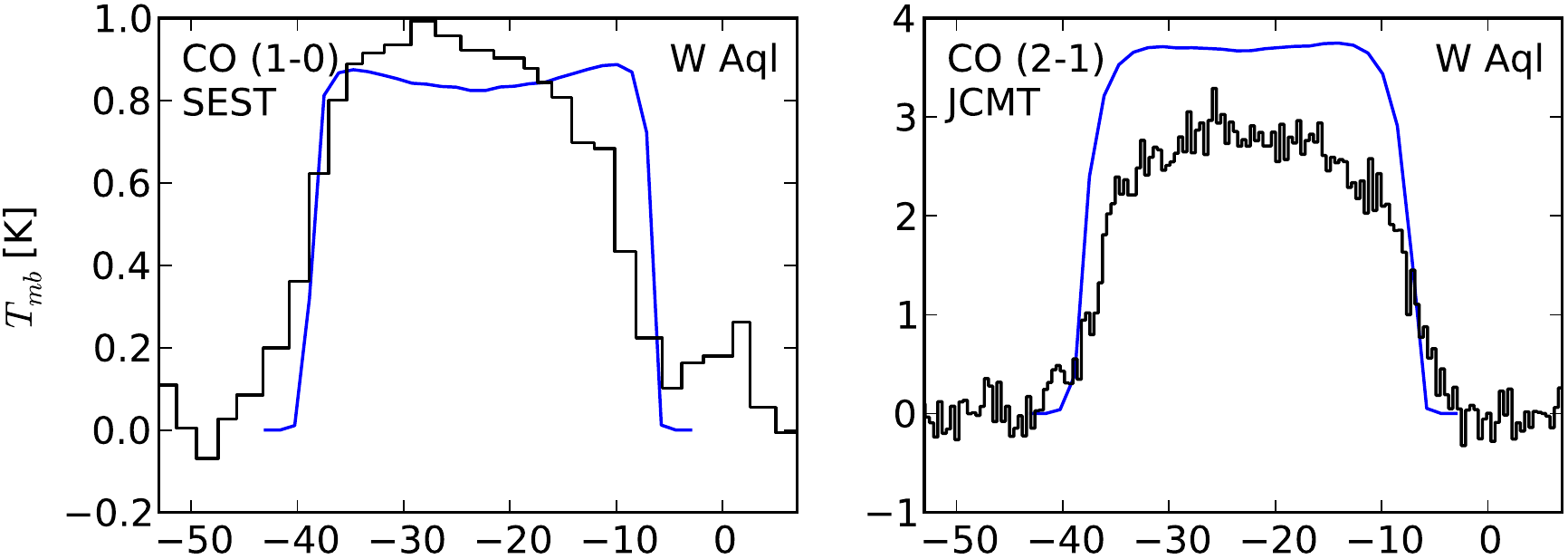}
   \includegraphics[width=8.8cm]{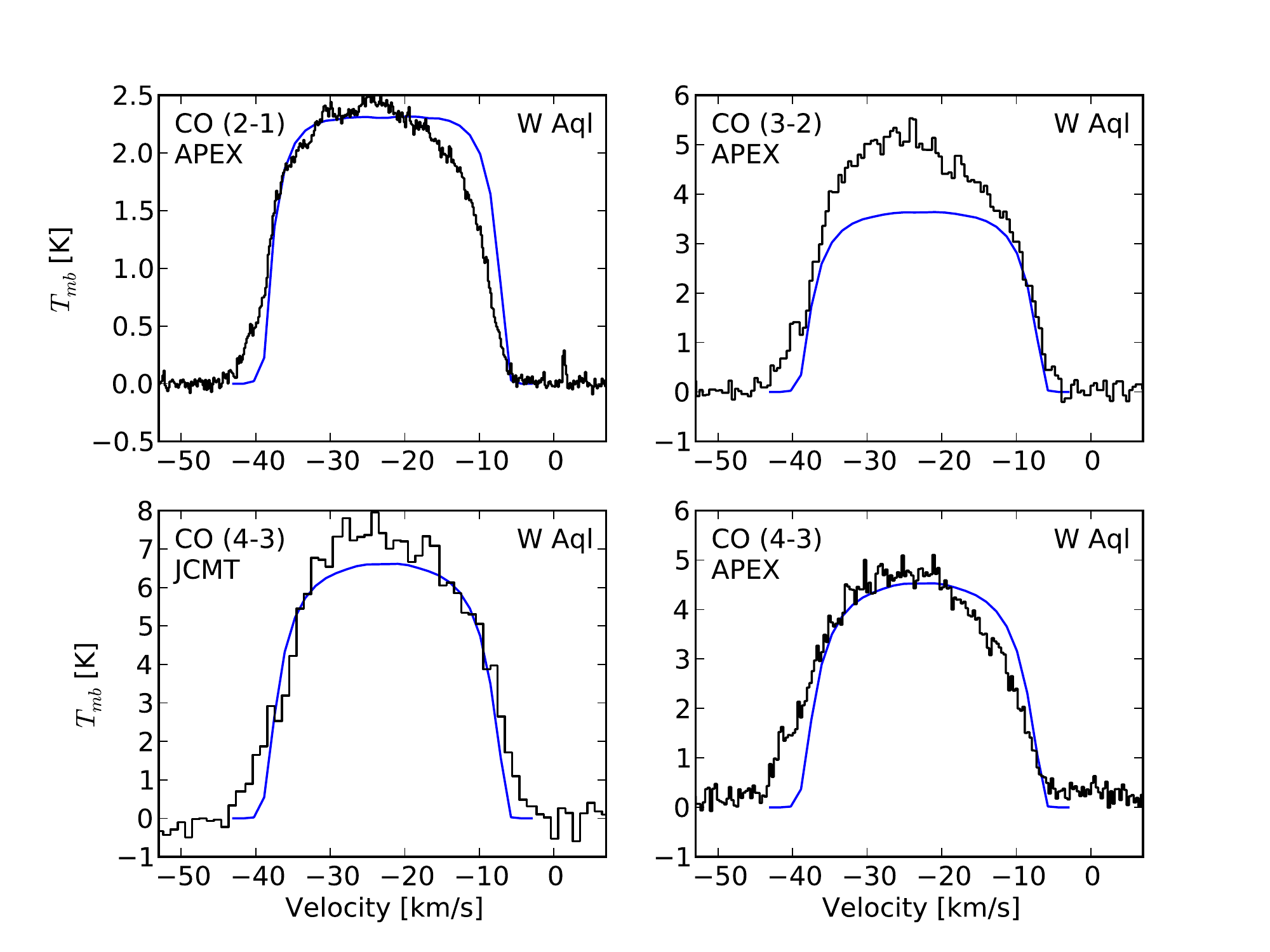}
   \includegraphics[width=8.8cm]{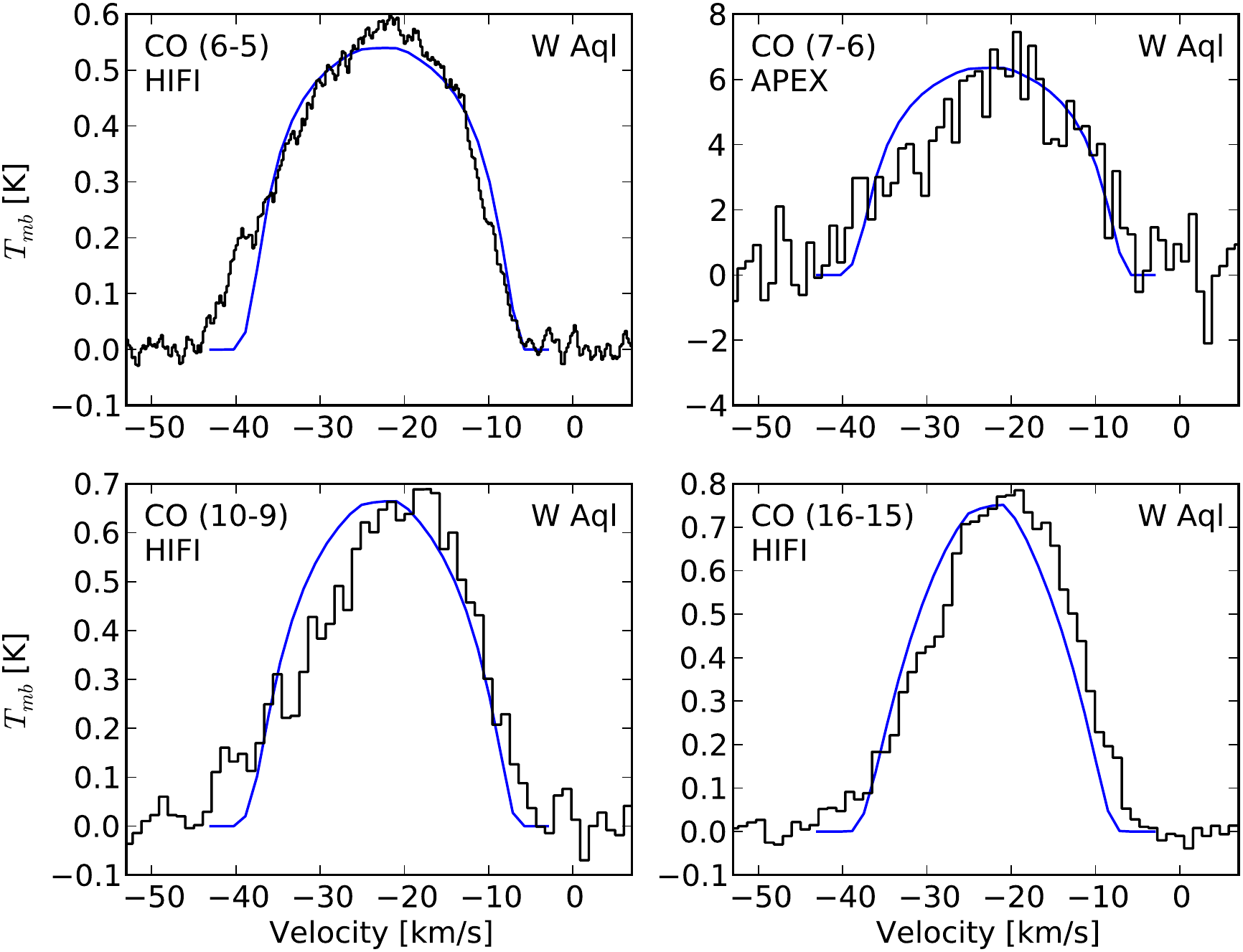}
      \caption{\up{12}CO model line profiles (solid blue lines) and observed data (black histograms). {Model parameters are listed in Table \ref{linemod}.}}
         \label{colines}
   \end{figure}

   \begin{figure}[tb]
   \centering
   \includegraphics[width=8.8cm]{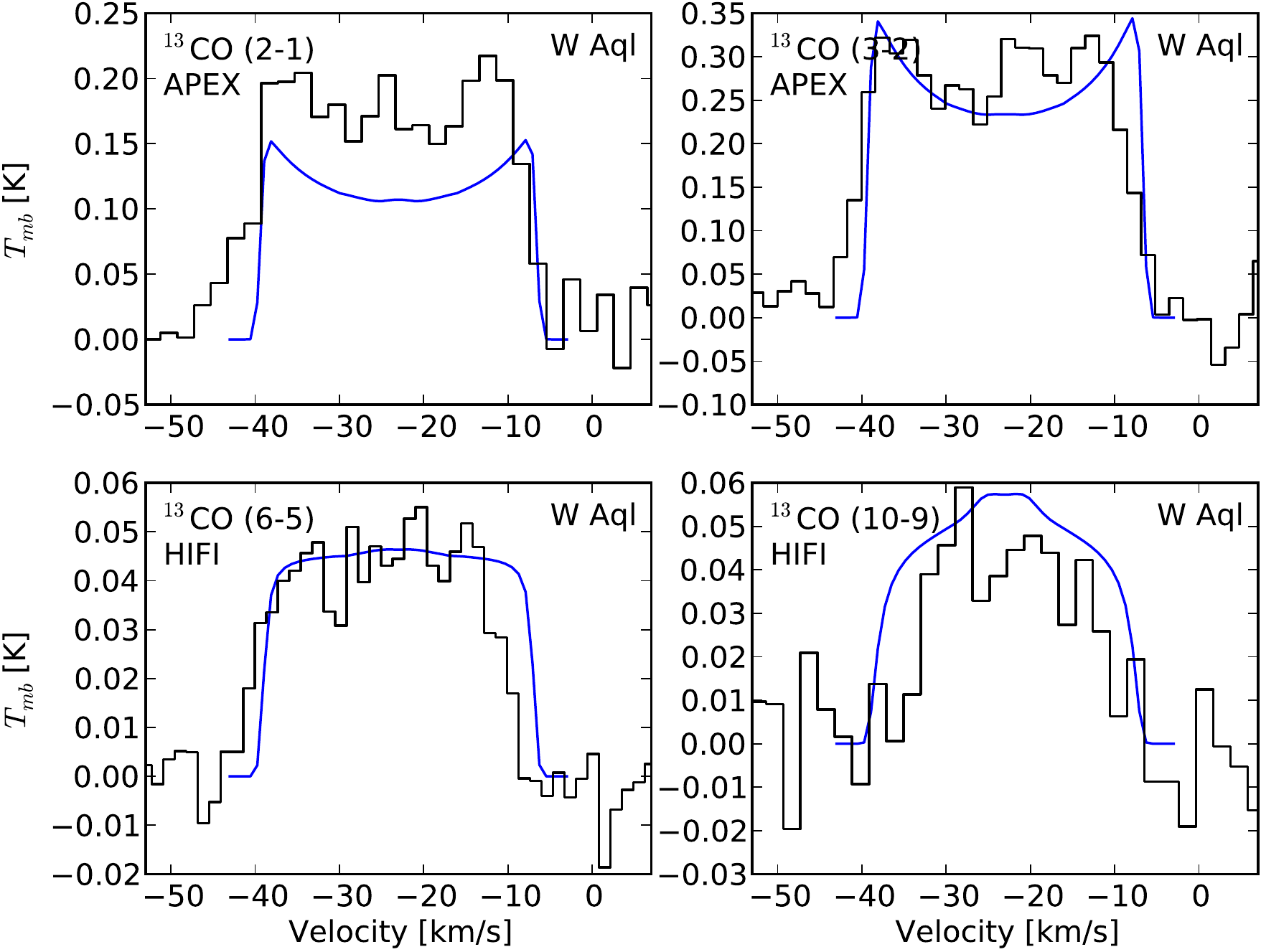}
      \caption{\up{13}CO model line profiles (solid blue lines) and observed data (black histograms). {Model parameters are listed in Table \ref{linemod}.}}
         \label{13colines}
   \end{figure}

The CO line radiative transfer modelling is based on the results of the dynamical and SED modelling for the stellar and dust properties of W Aql (in particular, the dust radiation field). A Monte-Carlo program (MCP), previously described in \citet{Schoier2001}, \citet{Schoier2002}, and \citet{Ramstedt2008}, is used to determine the molecular excitation. This is a well-tested, non-LTE and non-local, spectral line radiative transfer code, which has been benchmarked to high accuracy against other non-LTE line radiative transfer codes \citep{van-Zadelhoff2002}.

The gas energy balance, used in the MCP code when modelling the CO lines, includes heating due to dust and gas collisions and photoelectric heating due to the interstellar radiation field. The cooling processes are adiabatic cooling, \h2 line cooling, \h2O line cooling, CO line cooling, and cooling due to non-collisional interactions between dust and gas. The latter three can cause heating in certain circumstances. The heating and cooling processes are discussed in \citet{Schoier2001} and \citet{Schoier2011} in detail. The derived gas temperature structure for W Aql, which is consistent with the CO line intensities, is shown in Fig. \ref{temp}. { Line cooling from \h2O is discussed in \sec{watercool}. In Fig. \ref{heatcool}, we show all the contributions of the different heating and cooling terms including the \h2O line cooling that comes out of our \h2O model.}

In our CO excitation analysis, we included radiative transitions between rotational energy levels up to $J=40$ for the vibrational ground state and the first vibrationally excited state. For CO the energy levels, transition frequencies, and Einstein A coefficients were taken from \citet{Chandra1996}. Collisional rates, included for the ground vibrational state, are taken from \citet{Yang2010}, who calculated them separately for CO collisions with para- and ortho-\h2. The rates were weighed together assuming an ortho-/para-\h2 ratio of 3 by \citet{Schoier2011}. They cover temperatures from 2 to 3000 K. The same range of transitions and rates is used for \up{12}CO and \up{13}CO. { Collisional transitions within the vibrationally excited state and between the ground state and the vibrationally excited state can be ignored because of the fast radiative decay from the vibrationally excited state to the ground state.}

\begin{figure}[tb]
   \centering
   \includegraphics[width=8.8cm]{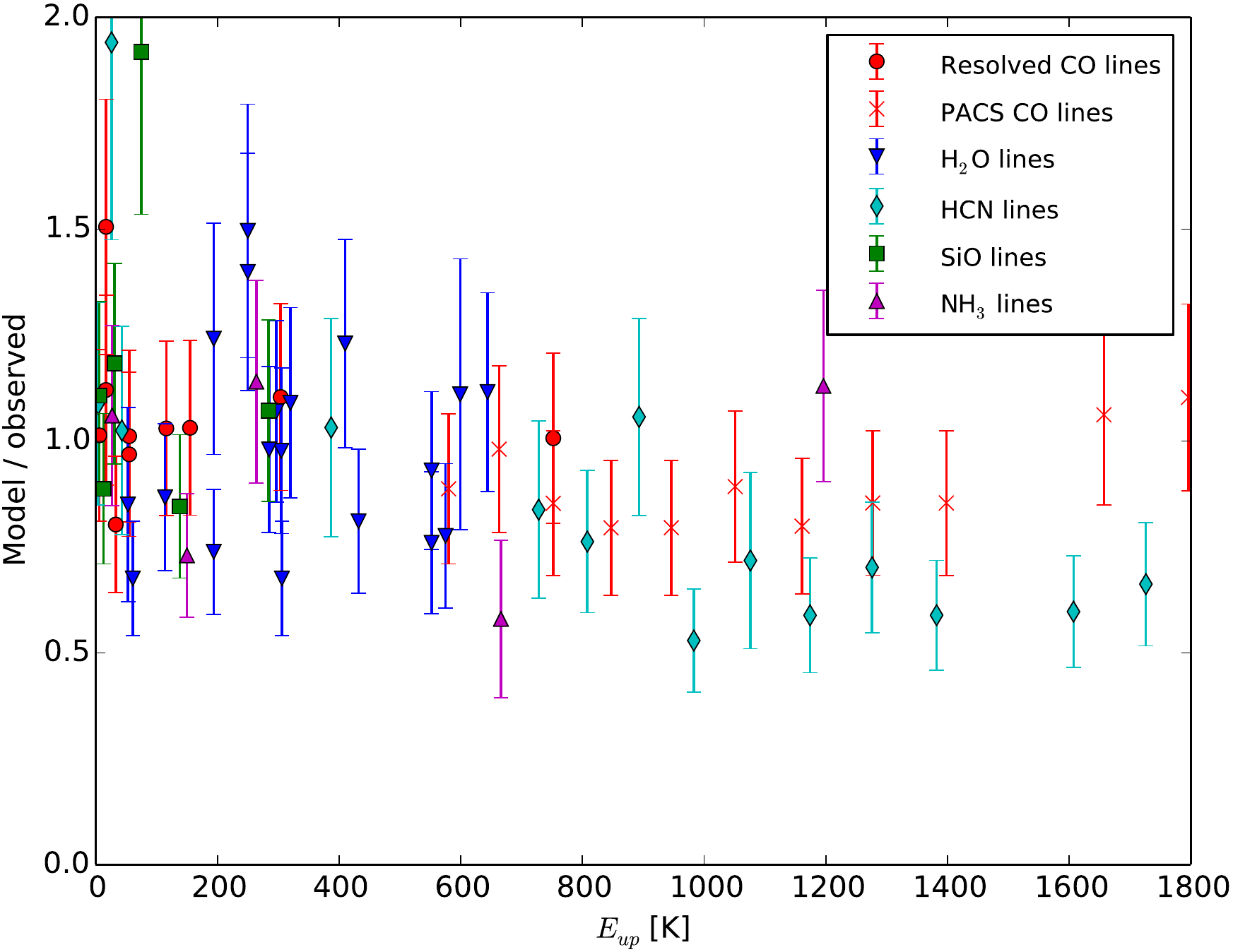}
      \caption{Model line intensities (resolved lines) or total fluxes (PACS lines) divided by the corresponding observed intensities or fluxes. The horizontal axis gives the energy of the upper level of the transition. {Model parameters are listed in Table \ref{linemod}.}}
         \label{Jratios}
   \end{figure}

{ To find the best fit model, we used all the resolved CO lines and PACS lines up to $J=25\rightarrow24$. Some discrepancies between measured line strengths in the HIFI and PACS spectroscopic data have been shown to exist for several sources observed with Herschel\footnote{http://hipecommunity.wikispaces.com/Herschel+spectral+cross+ calibration+webinar+1} with a bulk PACS/HIFI discrepancy in the \up{12}CO $16\rightarrow15$ line of 30\%. In our case, this is the only overlapping line between HIFI and PACS and our discrepancy is about 15\% (which, as can be see in Fig. \ref{Jratios}, is within the errors). Above $J=25\rightarrow24$, the lines detected in PACS have upper energy levels above 2000 K, which are believed to originate, at least in part, from within the dust condensation radius, which is outside of the scope of our model. These lines are hence excluded from our fitting procedure.
 It should be noted that for other molecules with PACS lines, all available lines were used in the $\chi^2$ analysis.
We show the results from the PACS CO lines in Table \ref{pacslines}, and include transitions up to $J=25\rightarrow24$ in Fig. \ref{Jratios} and in our $\chi^2$ value after the best fit was found.  

}

The size of the CO envelope is calculated in the MCP code from equations described in \cite{Schoier2001}, following the work of \cite{Mamon1988}. The initial CO abundance relative to \h2 at the inner radius is assumed to be $f_\mathrm{CO} = 6\e{-4}$, {adopted as the mean between common M star and C star values}. Using this formulation, we found the radius at which the abundance of CO has halved due to photodissociation to be $R_{1/2} = 1.1\e{17}$ cm, and the parameter which determines the CO radial abundance function to be $\alpha = 2.52$ \citep[for details see][]{Schoier2001} for the mass-loss rate of the best-fit CO model.

The only remaining parameters in the CO line fitting are the gas mass-loss rate and the gas-to-dust mass ratio, which enters through the gas-dust heating in the energy balance equation. { The parameter $\beta$ in the radial gas velocity profile (Eq.~\ref{veleq}) is also a free parameter in the fitting of the CO (and the \h2O) lines. It is however, only adjusted initially, before the best-fit solution is found, to get the best fit to the line shapes. It appears that the gas acceleration is slower ($\beta$\,=\,2) than predicted by the dust-driven wind model we employed ($\beta$\,=\,0.7).}

The results of our \up{12}CO line modelling are shown in Fig. \ref{colines}. Using the $\chi_\mathrm{red}^2$ statistic to find the best-fit model, we found a mass-loss rate of $\dot{M} = 4.0\e{-6} \spy$, and a dust-to-gas mass ratio $\Psi = 5\e{-3}$, a factor of 2.5 higher than found from the dynamical modelling. This result has $\chi^2_\mathrm{red} = 0.69$. { For each line, the ratio between the modelled line intensity (resolved lines) or flux (PACS lines) is included in Tables \ref{hifimol} (HIFI lines), \ref{pacslines} (PACS lines), and \ref{otherlines} (ground-based lines). The ratios are also plotted against the energy of the upper level of the transition in Fig. \ref{Jratios}. As can be seen, a good fit to the CO line data is obtained for a broad range of excitation energies. We estimate the uncertainty of the mass-loss rate to be of the order of three, when also taking into account uncertainties in the circumstellar model \citep[for a discussion on uncertainties in our modelling, see][]{Ramstedt2008}.}

A noticeable feature in the resolved \up{12}CO line profiles, which is absent in the model line profiles, is excess emission on the blue-shifted side. This will be discussed further in \sec{sec:asymmetry}.

Using the same envelope parameters as for \up{12}CO, and adjusting only the abundance to fit our \up{13}CO observations, we found $f_{^{13}\mathrm{CO}} = (2.1 \pm 0.5)\e{-5}$ with $\chi^2_\mathrm{red} = 2.0$ (the error here, and in the other estimated molecular abundances, is statistical within the adopted circumstellar model). The large $\chi^2$ is mostly due to the poor fit of the $J=2\rightarrow1$ line. The resulting \up{12}CO/\up{13}CO ratio $= 29 \pm 7$ is in agreement with the mean value \cite{Wallerstein2011} found for S stars (see also Ramstedt \& Olofsson, 2014, in press). Our model results are plotted with the data in Fig. \ref{13colines}.

\begin{table}
\caption{Parameters of W Aql and its CSE obtained from molecular line emission modelling.}
\label{linemod}      
\centering                          
\begin{tabular}{l r}        
\hline\hline                 
Mass-loss rate & $4.0\e{-6}\spy$\\
$\beta$ (see eq \ref{veleq}) & 2.0\\
Drift velocity & $9.5 \;\kms$\\
Dust to gas ratio, $\Psi$ & $5\e{-3}$\\
\hline                                   
\end{tabular}
\tablefoot{Descriptions of molecular abundance profiles are given in Table \ref{molres}.}
\end{table}

{ The low-$J$ \up{12}CO and \up{13}CO model lines are double-peaked (especially \up{13}CO), while the observed lines are round- or flat-topped. As noted by \cite{Olofsson2002} this is not an uncommon phenomenon among M stars. Possible ways to remedy this discrepancy are either to decrease the size of the molecular envelope or to increase the distance to the source. \cite{Olofsson2002} showed that, with suitable adjustments of mass-loss rate and gas temperature structure, it is indeed possible to fit the lower-$J$ lines much better with both these modifications. However, in our case, with the large number of higher-$J$ lines as further constraints, it has not been possible to obtain significantly better fits to the low-$J$ lines without also significantly increasing the $\chi^2_\mathrm{red}$ value of the overall fit with either of these modifications. We conclude that uncertainties in the gas temperature structure (perhaps due to a combination of dust properties), the possible non-spherical nature of the CSE, and/or the small-scale structure of the circumstellar gas (clumping) may provide possible explanations for the poor fit to the lower-J CO data.}

\subsection{H$_2$O line modelling}\label{water}

The radiative transfer modelling of the \h2O line emission uses a different code, since the MCP method used for CO and our other molecules is not capable of coping with the high optical depths encountered when modelling \h2O. Instead, an Accelerated Lambda Iteration (ALI) method is used, which is fully capable of dealing with high optical depths and takes maser action into account properly. Our ALI code has been previously implemented by \citet{Maercker2008,Maercker2009} \citep[including benchmarking against test cases in][]{van-Zadelhoff2002} and \citet{Schoier2011}, and is based on the ALI scheme described in \cite{Rybicki1991}. {We have also tested ALI against MCP solutions for CO line radiative transfer for several AGB CSE test cases and obtained results that agreed within a few percent. (The ALI code could, in principle, be used for the CO modelling but is not as it does include a solution of the energy balance equation.) Inputs such as mass-loss rate, gas temperature distribution, dust radiation field, etc. are taken from the SED and CO line modelling.}

We performed the radiative transfer analysis for ortho- and para-\h2O separately. { We included 45 rotational energy levels for each of the ground vibrational state, the first excited bending mode, $\nu_2=1$ at $6.3\um$, and the asymmetric stretching mode, $\nu_3=1$ at $2.7\um$.
The energy levels and radiative rates were obtained using the HITRAN database \citep{Rothman2009}. 
The collisional rates were taken from \citet{Faure2007} and cover eleven temperatures from 20 to 2000~K.} The collisional rates at lower temperatures are extrapolated following
\begin{equation}
c_{ul}(T) = c_{ul}(T_\mathrm{low}) \sqrt{\frac{T}{T_\mathrm{low}}}
\end{equation}
where $c_{ul}(T_\mathrm{low})$ is the collisional de-excitation rate between levels $u \rightarrow l$ at the lowest temperature for which collisional data are available.  

We assumed that the abundance distribution follows a Gaussian of the form 
\begin{equation}\label{abundance}
f = f_0 \exp\left( -\left( \frac{r}{R_e} \right)^2\right)
\end{equation}
where $f_0$ is the abundance at the inner radius, and $R_e$ is the $e$-folding radius, the radius at which the abundance has dropped by a factor of $1/e$. { This is a good first-order approximation to the abundance distribution of a species which is believed to be formed close to the star and eventually becomes photo dissociated by the interstellar radiation field.} For \h2O we take the $e$-folding radius as the photodissociation radius predicted by \citet{Netzer1987}, which \citet{Maercker2009} found to be consistent with their modelling of circumstellar H$_2$O line data using the same ALI method,

\begin{equation}
R_{\mathrm{H}_2\mathrm{O}} = 5.4\e{16}\left( \frac{\dot{M}}{10^{-5} \spy}\right)^{0.7} \left( \frac{\upsilon_\infty}{\kms} \right)^{-0.4} \;\mathrm{cm}
\end{equation}
This gives $R_{\mathrm{H}_2\mathrm{O}} = 9.3\e{15}$ cm for the estimated mass-loss rate. We found that this value resulted in a model that fits our data reasonably well.

\begin{figure}[tb]
   \centering
   \includegraphics[width=8.8cm]{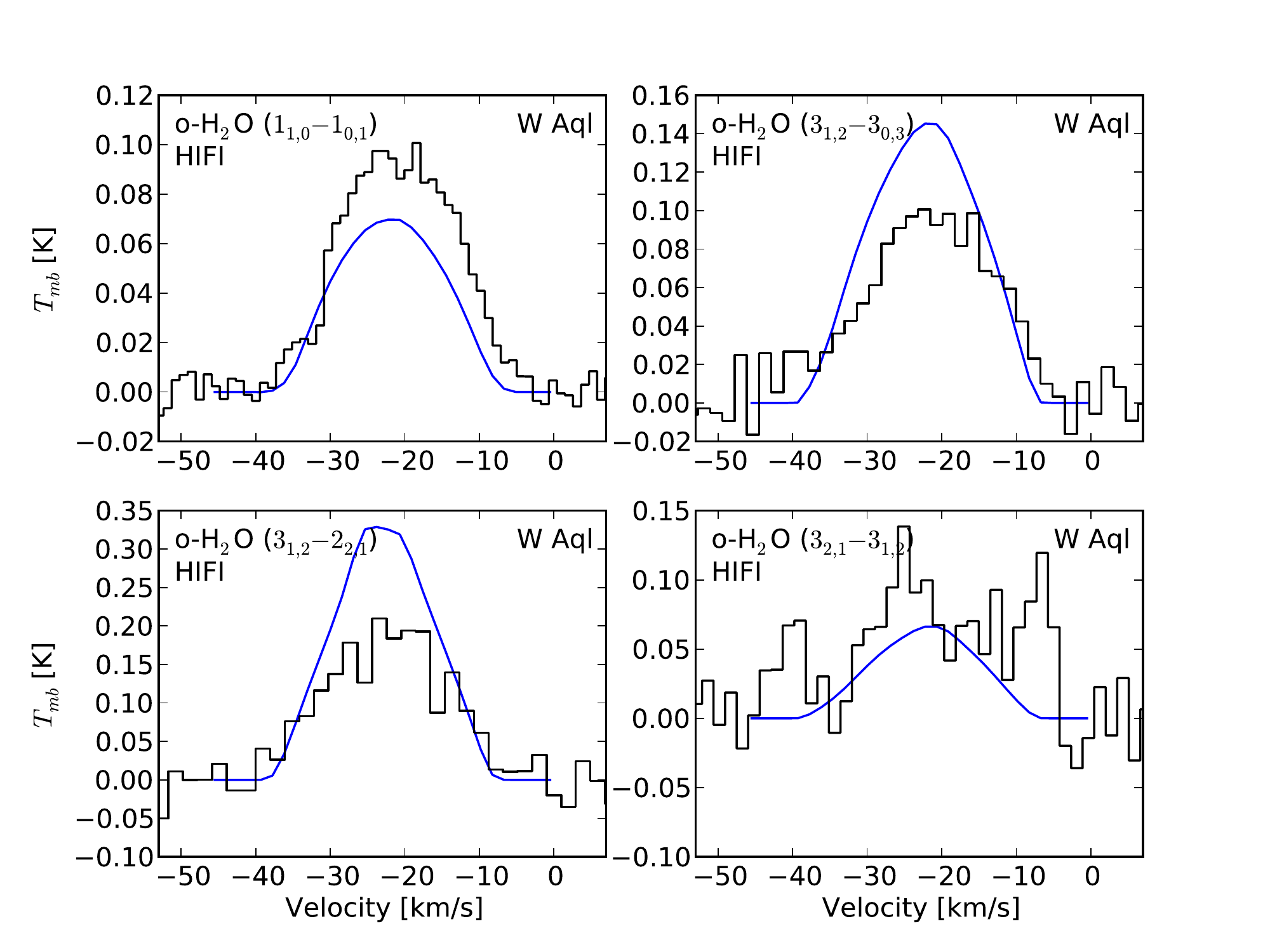}
   \includegraphics[width=4.4cm]{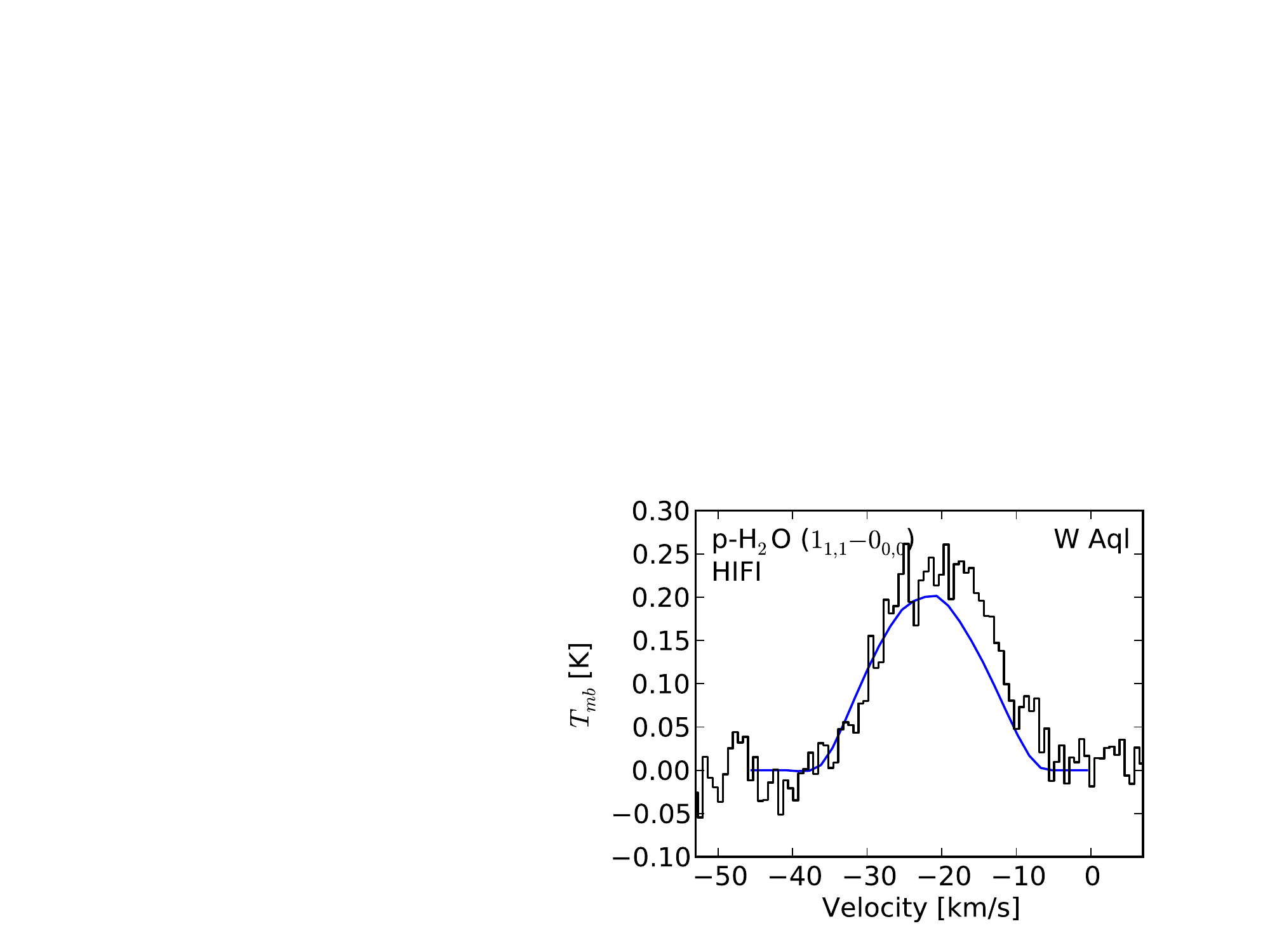}
      \caption{\h2O model line profiles (solid blue lines) and observed data (black histograms). {Model parameters are listed in Table \ref{linemod}.}}
         \label{waterlines}
   \end{figure}

In the case of H$_2$O, only the inner abundance $f_0$ is varied in the modelling. For ortho-\h2O we found an abundance of $f_{\mathrm{o-H}_2\mathrm{O}} = (1.0\pm 0.3)\e{-5}$ with $\chi^2_\mathrm{red} = 1.9$. 
For para-\h2O we found $f_{\mathrm{p-H}_2\mathrm{O}} = (4.5 \pm 2.0)\e{-6}$ with $\chi^2_\mathrm{red} = 0.46$. These results give an ortho-/para-\h2O ratio $= 2.2 \pm 1.2$, and a total \h2O abundance of $f_{\mathrm{H}_2\mathrm{O}} = (1.5\pm0.4)\e{-5}$. The results of our modelling for the HIFI lines are plotted with the data in Fig. \ref{waterlines}. {The ratios between the model and the observed fluxes for the PACS data are listed in Table \ref{pacslines} and plotted in Fig. \ref{Jratios}.}

\subsubsection{\h2O line cooling}\label{watercool}

Our ALI code outputs an \h2O line cooling file with the radial dependence of the rate based on the line cooling formula described in \cite{Sahai1990}
\begin{equation}\label{cool}
C_{\mathrm{H}_2\mathrm{O}}(r) = \sum_{l, u>l} \Delta E_{ul} k_B \left( n_l c_{lu} - n_u c_{ul} \right)
\end{equation}
where $n_l$ is the population of level $l$, $c_{lu}$ the collisional excitation rate from $l$ to $u$, $k_B$ Boltzmann's constant and $\Delta E_{ul}$ the difference in energy between levels $u$ and $l$. This is the same formula used in the MCP code for CO line cooling.

{ Separate cooling functions were obtained for ortho- and para-\h2O and they were included as part of the energy balance in the CO line modelling. This was achieved by finding a CO model, using MCP, which fits the data without \h2O line cooling. The CO model gas temperature profile was used as an input to the ALI code to find an \h2O model which fits the \h2O data. The resulting cooling function was then included in the CO line modelling, which was once again adjusted to best fit the observed lines.
After a few iterations, a consistent model between CO and \h2O was found. The total \h2O line cooling contribution can be seen in Fig.~\ref{heatcool}.}

\subsection{Analysis of the other molecular line emissions}

The radiative transfer analysis for the remaining molecules, other than NH\down{3}, was done using the same MCP code as for CO. Mass-loss rate, gas temperature distribution, dust radiation field, etc. are taken from the SED and CO line modelling. What remains to be determined for the other molecules are the abundances. We use \eq{abundance} to describe the abundance distributions of the remaining molecules, except CN.

\subsubsection{HCN line modelling}\label{HCN}

In our excitation analysis of HCN, we included the vibrational ground state and first excited states for each of the three vibrational modes: the CH stretching mode (100) at 3\,$\um$, the bending mode (010) at 14\,$\um$, and the CN stretching mode (001) at 5\,$\um$. We included rotational transitions in each of these states up to $J$\,=\,29 with hyperfine splitting included for $J$\,=\,1 in the ground vibrational state and for the lowest rotational state in each excited vibrational state. For the $14\um$ transitions $l$-type doubling --- splitting the (010) vibrational level into (01$^{1c}$0) and (01$^{1d}$0) --- was included \citep{Schoier2011}.

The collisional rates were taken from the \citet{Dumouchel2010} rates calculated for HCN and He, scaled by 1.363 to represent collisional rates between HCN and \h2. Our collisional data included collisions for the ground vibrational state up to the $J$\,=\,25 rotational level and covers 25 collision temperatures between 5 and 500~K. No collisional rates coupling vibrational states were included as these are insignificant compared with radiative rates between the same levels. The collisional rates for the hyperfine levels were calculated assuming hyperfine statistical equilibrium \citep{Keto2010}
\begin{equation}
c_{J,K\rightarrow J'K'} = \frac{g(J,K)}{g(J)}c_{J\rightarrow J'}\,,
\end{equation}
where $g(J)$ is the statistical weight for the level without hyperfine splitting, and $g(J,K)$ the weight for the $K$ hyperfine level. The collisional rates between hyperfine levels (for $\Delta J = 0$) were assumed to be 0. We assumed the same collisional cross-sections for H\up{13}CN and used the same number of energy levels and transitions in the H$^{13}$CN excitation analysis. 

We calculated the HCN $e$-folding radius using the relation found in \cite{Schoier2013}
\begin{equation}
\log R_\mathrm{HCN} = 19.9 +0.55 \log \left( \frac{\dot{M}}{\upsilon_\infty}\right)
\end{equation}
where $R_\mathrm{HCN}$ is in cm, $\dot{M}$ in$\spy$, and $\upsilon_\infty$ in $\kms$. This gives $R_\mathrm{HCN} = 1.8\e{16}$ cm for the estimated mass-loss rate, { a value that we use for both HCN and H\up{13}CN}. We found that this value resulted in a model that fits our data well. 

\begin{figure}[tb]
   \centering
   \includegraphics[width=8.8cm]{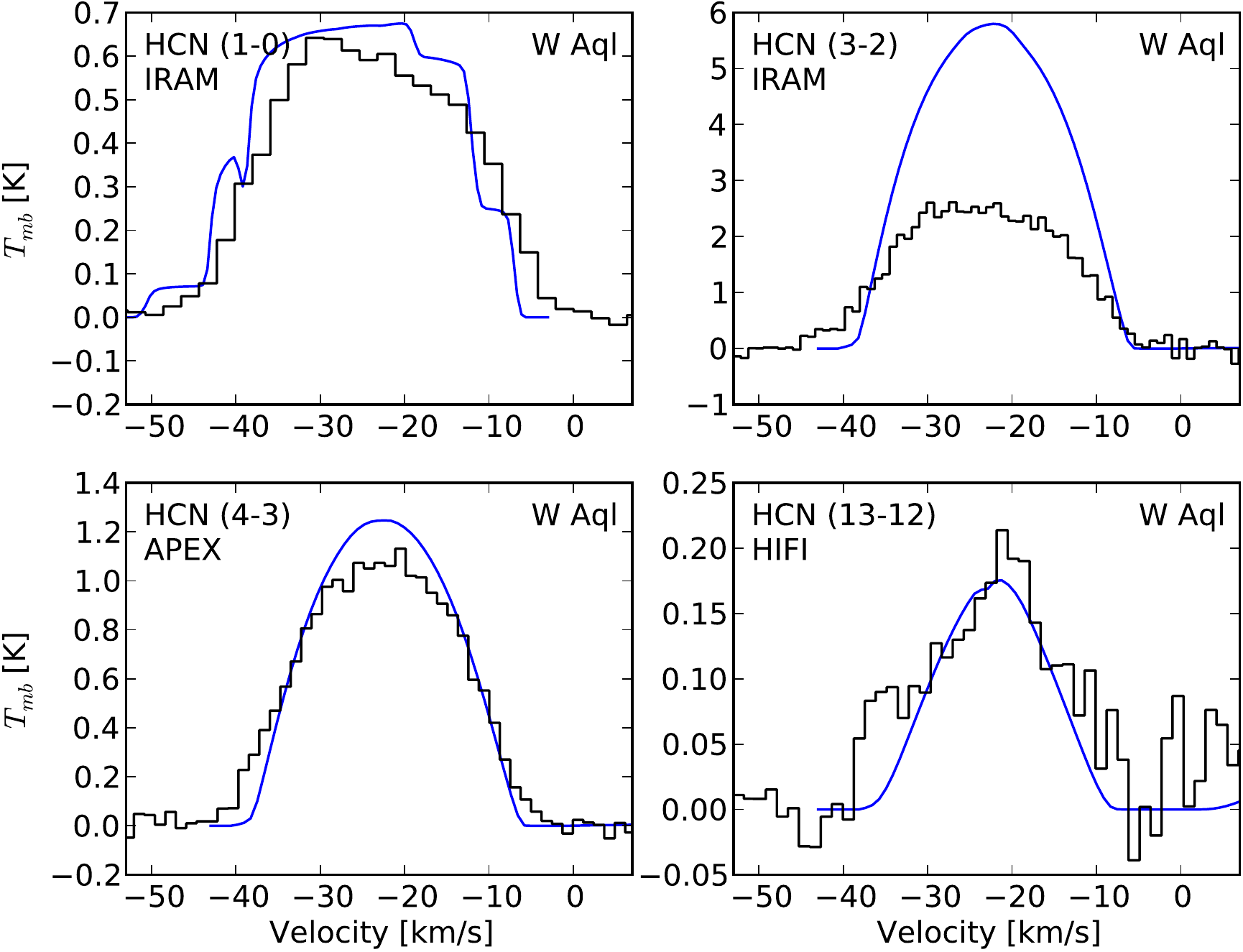}
   \includegraphics[width=4.4cm]{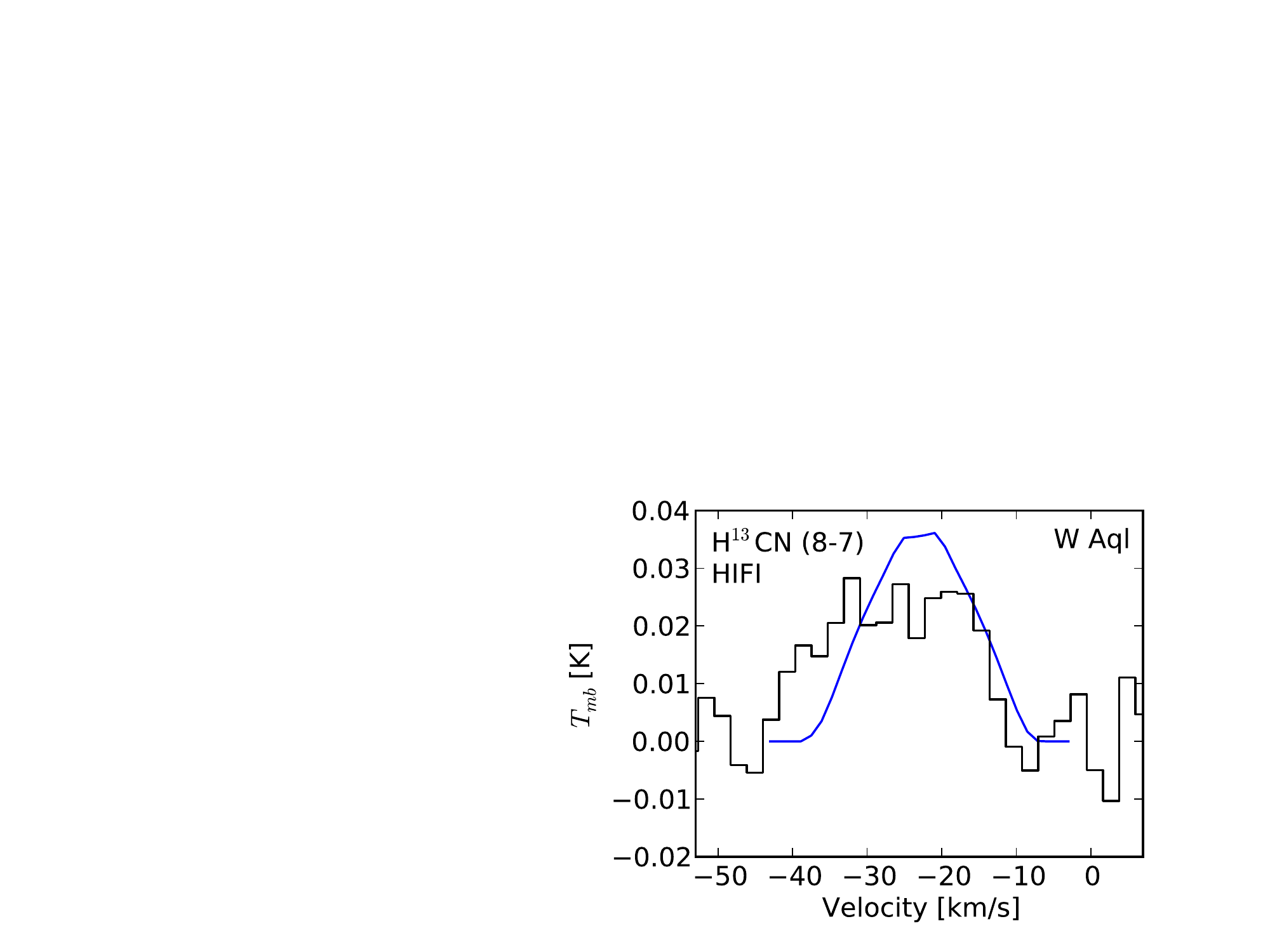}
      \caption{H\up{12}CN and H\up{13}CN model line profiles (solid blue lines) and observed data (black histograms). {Model parameters are listed in Table \ref{linemod}.} Note that the hyperfine structure of the \mbox{$J=1\rightarrow0$} line is reasonably well fitted.}
         \label{hcnlines}
   \end{figure}

For HCN we found an abundance of $f_\mathrm{HCN} = (3.1 \pm 0.1)\e{-6}$. This result has $\chi^2_\mathrm{red} = 3.3$. Plotting the resolved lines in Fig.~\ref{hcnlines}, it is evident there is a particular discrepancy in the $J=3\rightarrow2$ line. { It is possible that the discrepancy in this line could be due to dependence on the pulsation phase, particularly as most lines were observed at different times, apart from the group of PACS lines. The same line was also observed with the Heinrich Hertz Submillimeter Telescope by \cite{Bieging2000} with a comparable result when taking the different telescope parameters into account, which suggests an anomaly in the line itself, rather than in our individual observation.}

For H\up{13}CN we found $f_{\mathrm{H}^{13}\mathrm{CN}} = (2.8\pm0.8)\e{-7}$, which gives a ratio H\up{12}CN/H\up{13}CN $= 11 \pm 3.5$ that does not agree with the ratio derived from CO. However, with only one weak H\up{13}CN line in the analysis, the uncertainty in its abundance is significantly larger than the statistical value obtained from the uncertainty of a single line intensity. Hence, the result is not significant. The model results for HCN and H\up{13}CN are presented in Fig. \ref{hcnlines} together with the observed lines. The ratios between modelled and observed lines are plotted in Fig. \ref{Jratios} and listed in Tables \ref{hifimol}, \ref{pacslines} and \ref{otherlines}.

\subsubsection{SiO line modelling}

Our excitation analysis of SiO follows the same approach as HCN. Radiative transitions for the ground and first excited vibrational states up to rotational energy level $J$\,=\,40 are included. Energy levels, transition frequencies, and Einstein A coefficients within the first two vibrational states were taken from CDMS \citep{Muller2001, Muller2005}. The Einstein A coefficients for the ro-vibrational transitions were calculated using the electric dipole moment measured by \citet{Raymonda1970}. The collisional rates, included for the ground vibrational state, were taken from collisions with He by \citet{Dayou2006} scaled by 1.38 and with extrapolations calculated by \citet{Schoier2005}. This gave collisional rates covering temperatures in the range 10--2000 K and up to $J=40$. For $^{29}$SiO, we included the same energy levels in our excitation analysis and the same collisional rates as for $^{28}$SiO. 

We adopt the SiO $e$-folding radius formula determined by \cite{Gonzalez-Delgado2003}
\begin{equation}
\log R_{\mathrm{SiO}} = 19.2 + 0.48 \log \left(\frac{\dot{M}}{\upsilon_\infty}\right)
\end{equation}
where $R_{\mathrm{SiO}}$ is in cm, $\dot{M}$ in$\spy$, and $\upsilon_\infty$ in $\kms$. This gives $R_\mathrm{SiO} = 1.0\e{16}$ cm for the estimated mass-loss rate, { a value that we use for both SiO and \up{29}SiO}. We found that this value resulted in a model that fits our data well.

When constraining our model, we found that the SiO $J$\,=\,8$\rightarrow$7 model line consistently came out much brighter than the observed line. This remained true when we compared our models with the $J$\,=\,8$\rightarrow$7 observations in \citet{Ramstedt2008}, which were observed using both the JCMT and APEX. Including the $J$\,=\,8$\rightarrow$7 line we found $\chi^2_\mathrm{red} =4.6$, but excluding that line we found $\chi^2_\mathrm{red} =0.54$. { This discrepancy could be due to time variations in temperature and luminosity --- although this seems unlikely given the issues exists for unrelated observations taken at different times with different telescopes --- or could be the result of line overlaps as investigated by \citet{Pardo1998} and \citet{Herpin2000}.}

The abundances we found were $f_\mathrm{SiO} = (2.9\pm0.7) \e{-6}$ and $f_{^{29}\mathrm{SiO}} = (2.3\pm0.6)\e{-7}$. This gives a \up{28}SiO/\up{29}SiO abundance ratio of $13 \pm 6$, which is twice the ratio of $\sim 7$ found by \citet{Schoier2011} for $\chi$ Cyg. 

\begin{figure}[tb]
   \centering
   \includegraphics[width=8.8cm]{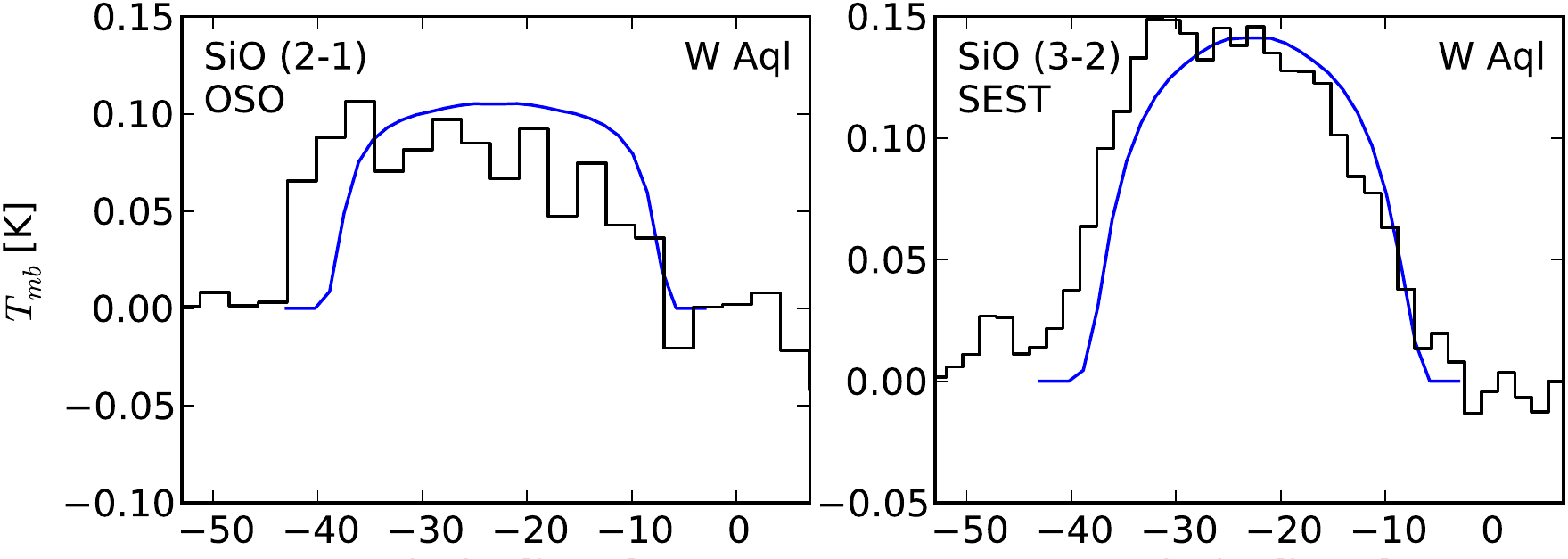}
   \includegraphics[width=8.8cm]{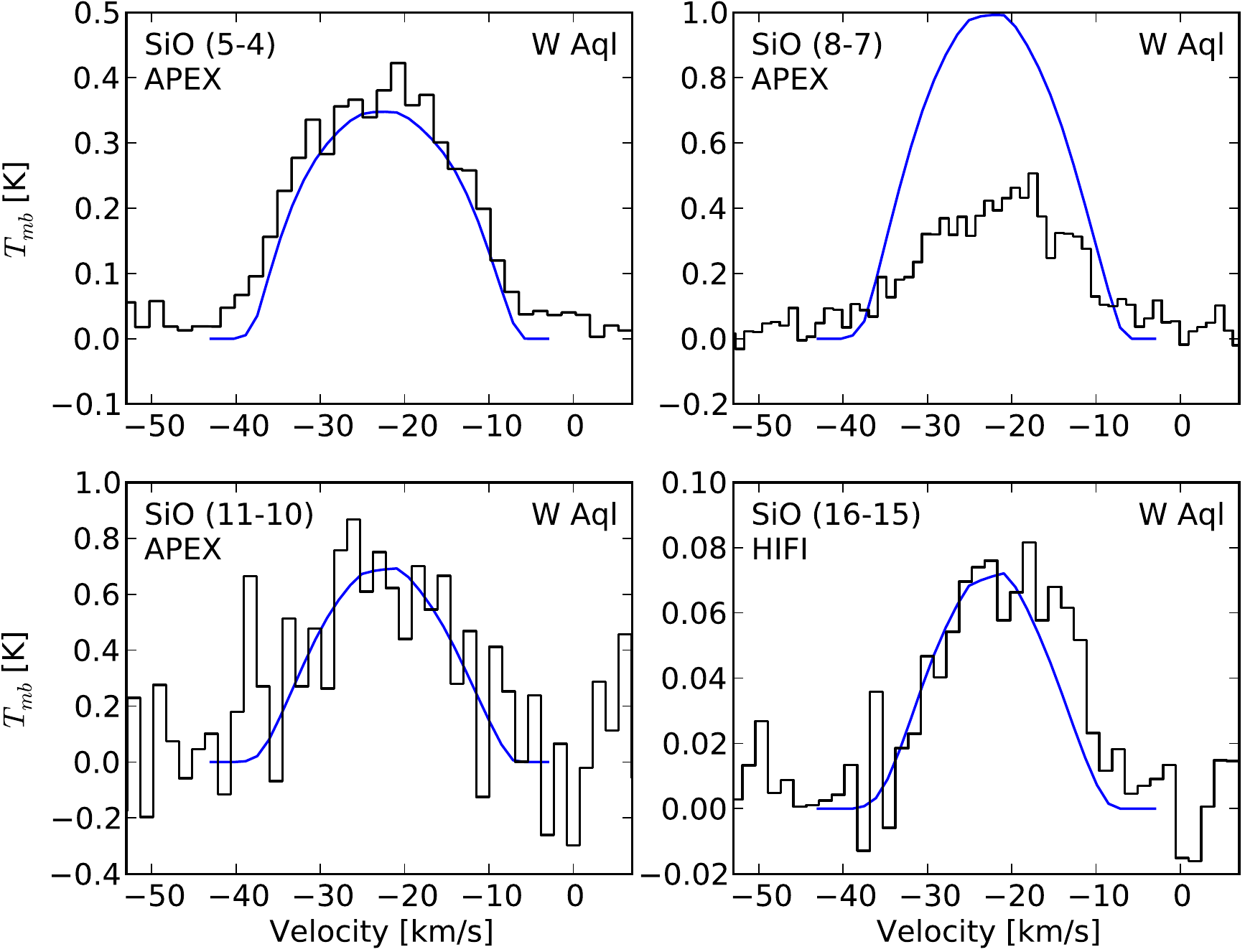}
   \includegraphics[width=4.4cm]{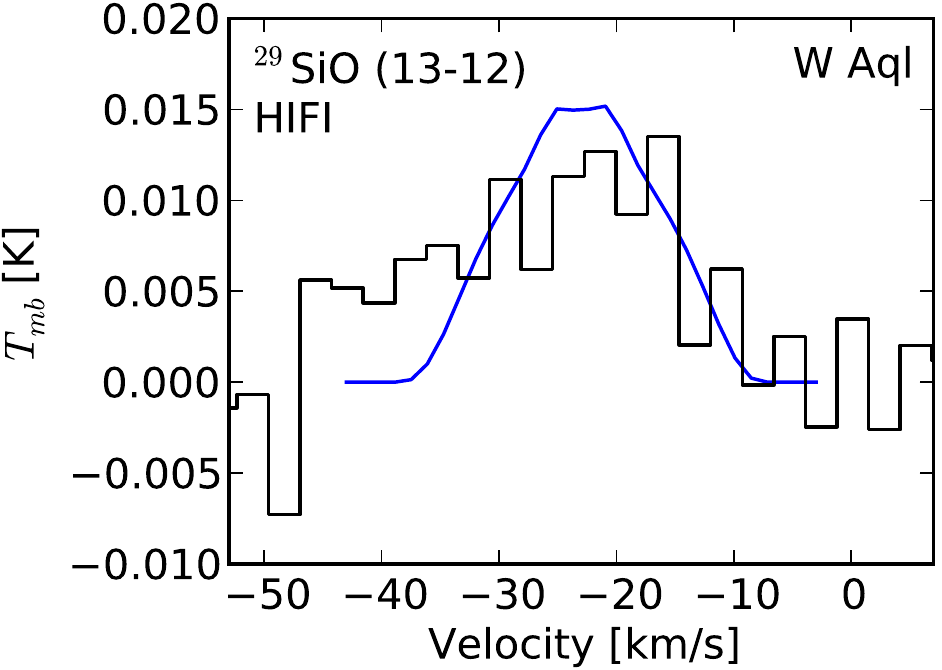}
      \caption{\up{28}SiO and \up{29}SiO model line profiles (solid blue lines) and observed data (black histograms). {Model parameters are listed in Table \ref{linemod}.}}
         \label{siolines}
   \end{figure}

However, our \up{29}SiO model, as well as that of \citet{Schoier2011}, is highly uncertain as it is based on only one observed \up{29}SiO line,  { and should not be taken as a definitive estimate of the \up{29}SiO abundance.} The model results and the observed data are shown in Fig. \ref{siolines}.

\subsubsection{NH$_3$ line modelling}

We model NH\down{3} using the ALI code, which copes better with the large number of energy levels and transitions we included in our analysis. Energies and radiative transitions for the ground and first excited vibrational states up to the $J=12$ rotational levels are included, as are the inversion transitions in that range (3188 transitions in total). The molecular data were obtained from \cite{Yurchenko2009}. The collisional rates for the ground vibrational state were taken from \cite{Danby1988} for collisions between NH\down{3} and para-\h2 with the combined ortho- and para-\h2 collisional rates adapted from BASECOL and temperatures ranging from 15 to 300 K. Above $J=5$ the adopted collisional rates were scaled from the radiative rates. { The transitions of the $\nu_1$ and $\nu_3$ modes are inherently 15--20 times weaker than those of $\nu_2$ and can most likely be safely ignored in the radiative transfer analysis. The transitions of the $\nu_4$ mode are several orders of magnitude weaker than those of $\nu_2$ and can safely be ignored. Therefore we  only include the ground and $\nu_2 = 1$ states in our analysis.Ó

Modelling of NH\down{3} by \citet{Menten2010} and \citet{Schoier2011} was based primarily on the ground state line observed by HIFI but in the case of \citet{Menten2010}, two inversion lines were also included. The \citet{Menten2010} study did not include the possibility of infrared pumping of the $1_0\rightarrow0_0$ line and found unexpectedly high abundances in the oxygen-rich stars they studied. \citet{Schoier2011} found that the inclusion of the first vibrationally excited state, $v_2=1$ the symmetric bending mode at 10 $\mic$, in the radiative transfer analysis lowered the derived fractional abundance of NH$_3$ by about an order of magnitude, due to IR pumping. However, \citet{Schoier2011} assumed that the NH\down{3} envelope was the same size as the \h2O envelope. With the inclusion of PACS NH\down{3} lines, some of which come from the inner part of the envelope, we are able to place better constraints on the size of the NH\down{3} envelope.

To find a model which agreed with both the PACS and HIFI observations, we had to significantly reduce the $e$-folding radius and increase the fractional abundance.}
Treating ortho- and para-NH\down{3} together, { we found a total fractional abundance $f_{\mathrm{NH}_3}=(1.7\pm1.0)\e{-5}$ and an e-folding radius $R_e = (1.0 \pm0.3)\e{15}$ cm with a $\chi^2_\mathrm{red} = 1.1$.} 

\begin{figure}[tb]
   \centering
   \includegraphics[width=4.4cm]{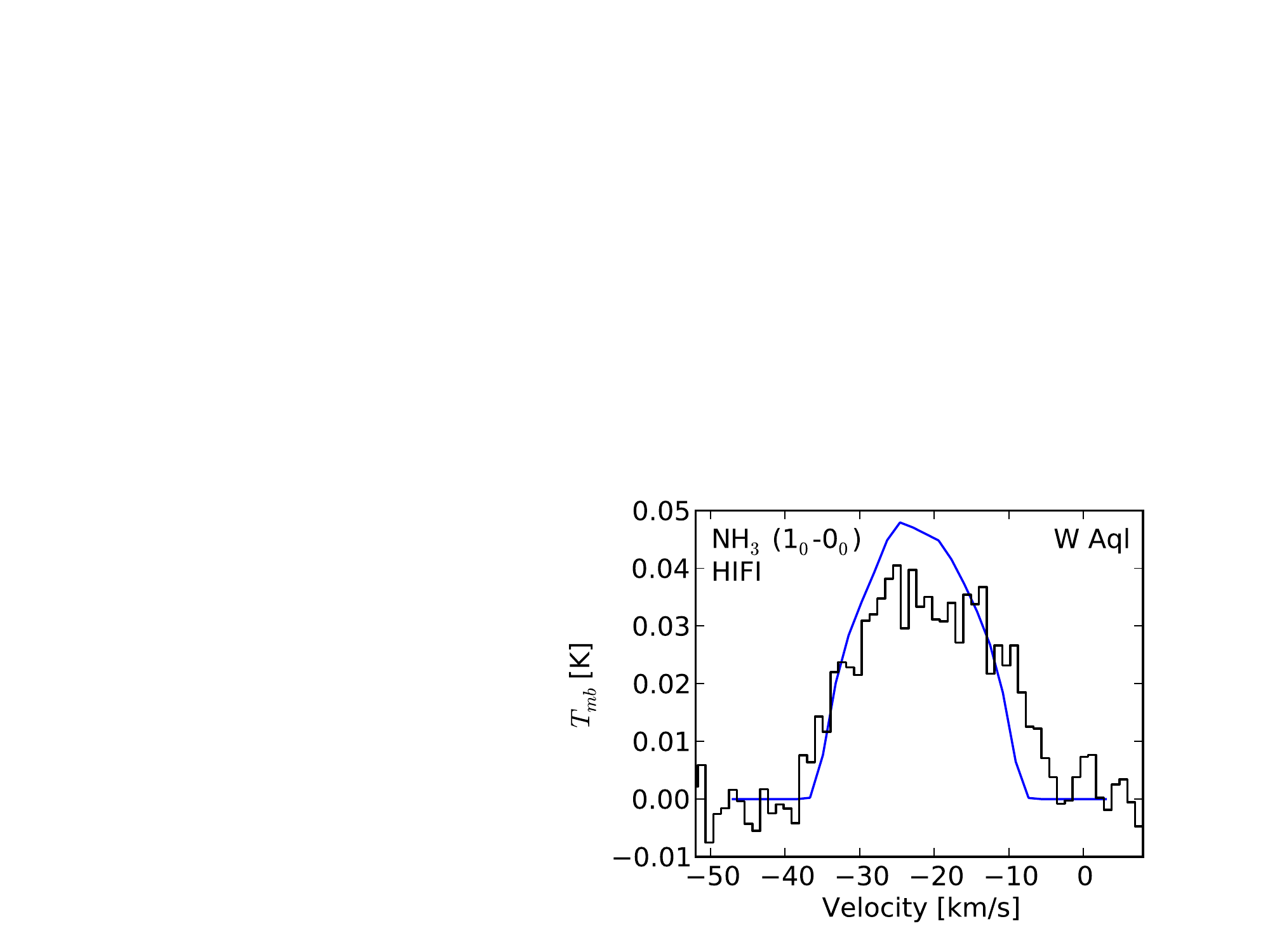}
      \caption{NH$_3$ model line (for the model including IR pumping; solid blue line) and the observed spectra (histogram). {Model parameters are listed in Table \ref{linemod}.}}
         \label{nh3lines}
   \end{figure}

The model result and the ground state line are shown in Fig. \ref{nh3lines} and the ratio of modelled to observed lines is shown in Fig. \ref{Jratios} and listed in Tables \ref{hifimol} and \ref{pacslines}. Also note the comparatively small size of the NH\down{3} envelope plotted with the other molecules in Fig. \ref{abundances}.

\subsubsection{CN line modelling}

CN was not detected in W Aql in the HIFISTARS project although the CN $(6_{13/2}\rightarrow 5_{11/2})$ and $(6_{11/2}\rightarrow 5_{9/2})$ lines fall in the frequency range of setting 17. However, the $N=2\rightarrow 1$ and $N=1 \rightarrow 0$ line groups were detected by \cite{Bachiller1997}. We estimated the CN abundance based on the \cite{Bachiller1997} data and the detection limit in our data.

{ Our excitation analysis included the spin-rotation structure, the hyperfine structure, and energy levels up to $N=11$ in the ground vibrational state. The energy levels, transition frequencies and Einstein coefficients were obtained from \cite{Klisch1995}. The collisional rates were adapted from CO rates by \citet{Schoier2011} and cover temperatures from 10 to 100 K (and are linearly extrapolated for higher temperatures).}

Usually CN is found in a shell (a photodissociation product of HCN) and not with a peak abundance centred on the star as we have assumed for the other molecules in this paper. \citet{Lindqvist2000} found for C stars that the position of the shell, and the difference between peak and central abundances, can vary significantly. On the other hand, \citet{Schoier2011} found that, for $\chi$ Cyg, a Gaussian distribution centred on the star gave a better fit to the observed data than a shell distribution. Our non-detection of the high-$J$ CN lines with HIFI suggests a low abundance in the inner region, with a higher abundance in the outer region as suggested by the \citeapos{Bachiller1997} detection, i.e., a shell-like distribution. We therefore adopted a CN distribution following
\begin{equation}\label{cnab}
f = f_\mathrm{CN} \exp \left( -4.6 \frac{(r-R_\mathrm{peak})^2}{R_w^2}\right)
\end{equation}
This is a first-order approximation of the abundance of a species formed through photodissociation. We used the HCN $e$-folding radius as the peak radius, $R_\mathrm{peak} = 1.8\e{16}$ cm, and fit $f_\mathrm{CN}$ and $R_w$, a measure of the shell width, to the available data, including constraints from the HIFI upper limits. $f_\mathrm{CN}$ is primarily constrained by the \cite{Bachiller1997} observations, while the HIFI detection limit allows us to put an upper limit on $R_w$.

   \begin{figure}[tb]
   \centering
   \includegraphics[width=8.8cm]{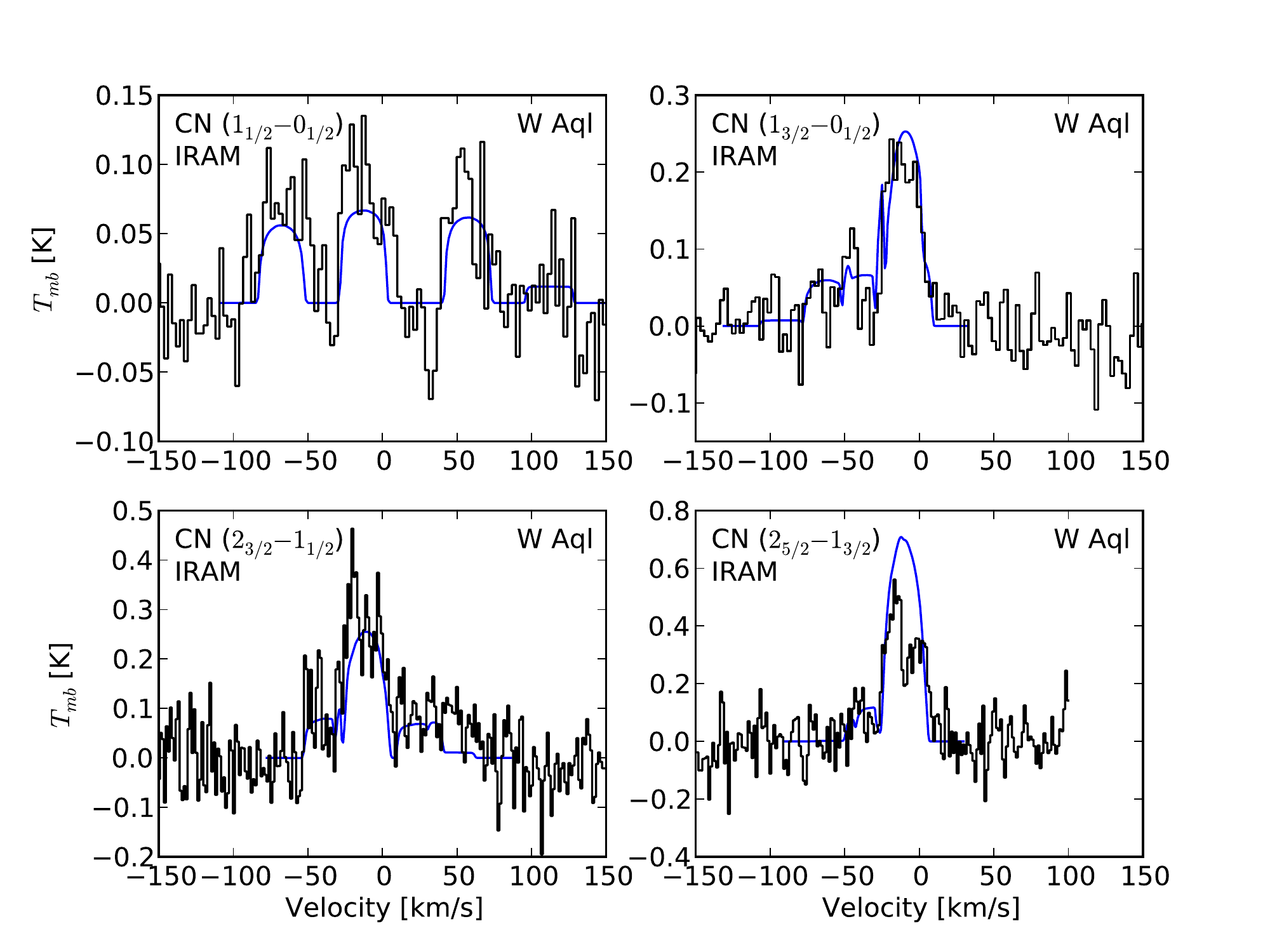}
      \caption{ CN model lines (solid blue line) and the observed spectrum (histogram). {Model parameters are listed in Table \ref{linemod}.}}
         \label{cnlines}
   \end{figure}

We found a peak CN abundance of $f_\mathrm{CN} = 5.7\e{-6}$ and a shell width $R_w = 3.0\e{16}$ cm. This is within the expected uncertainties of the \citeapos{Bachiller1997} results of $f_\mathrm{CN} = 1.2\e{-5}$ and $R_\mathrm{peak} = 2\e{16}$ cm (we cannot compare our $R_w$ result as their $r_i$ is defined differently). However, our abundance estimate is surprisingly high if CN is taken to be an HCN dissociation product, since our HCN abundance is $3.1\e{-6}$. {As can be seen in Fig.~\ref{Jratios}, our HCN model tends to underpredict the PACS lines and the best fit model is somewhat skewed by the anomalous $J=3\rightarrow2$ line. This could mean that we under-estimate the HCN abundance, although likely not by a factor of two. It is also possible that the lack of excited vibrational states in our excitation analysis contributed to an over-estimation of the CN abundances. A similar effect has been seen with NH\down{3} by \citet{Schoier2011}. Our CN model results are shown in Fig. \ref{cnlines} together with the data of \cite{Bachiller1997}.}

\section{Discussion}\label{sec:dis}

\begin{table}[tb]
\caption{Estimated molecular abundances in the CSE of W Aql.}
\label{molres}  
\centering   
\begin{tabular}{lcccc}
\hline\hline
Molecule & $f_0$ & $R_e$ [cm] & $\chi^2_\mathrm{red}$ & N \\
\hline
CO			& $6.0\e{-4}$	&$\;\,1.1\e{17}$:	& 0.69 & 21\\
$^{13}$CO	& $(2.1\pm0.5)\e{-5}$	&$\;\,1.1\e{17}$:	&2.0& 4\\
o-\h2O	& $(1.0\pm0.3)\e{-5}$	& $9.3\e{15}$	&1.9& 13\\
p-\h2O	& $(4.5\pm2.0)\e{-6}$	& $9.3\e{15}$	&\phantom{0}0.46& 6\\
HCN			& $(3.1\pm0.1)\e{-6}$ 	& $1.8\e{16}$ & 3.3& 14\\
H$^{13}$CN	& $(2.8\pm0.8)\e{-7}	$	& $1.8\e{16}$ & ...& 1\\
SiO			& $(2.9\pm0.7)\e{-6}$	& $1.0\e{16}$	& \phantom{0}0.54 & 6*\\
$^{29}$SiO	& $(2.3\pm0.6)\e{-7}$	& $1.0\e{16}$	&...& 1\\
\nh3	& $(1.7\pm1.0)\e{-5}$	& $(1.0\pm 0.3)\e{15}$ & 1.4& 4\\
CN			& $5.7\e{-6}$	& $\;1.8\e{16}$$^\dagger$& 1.3 & 4\\
\hline
\end{tabular}
\tablefoot{{ In general, the parameters listed are those of eq \ref{abundance}. However,} : indicates that the half-abundance radius, $R_{1/2}$, is listed; $^\dagger$ indicates that $R_\mathrm{peak}$, the radius of peak abundance, is listed (see eq \ref{cnab}). $N$ is the number of molecular lines included in the analysis. * indicates that one line was excluded from the $\chi^2$ analysis.}
\end{table}

   \begin{figure}[tb]
   \centering
   \includegraphics[width=8.8cm]{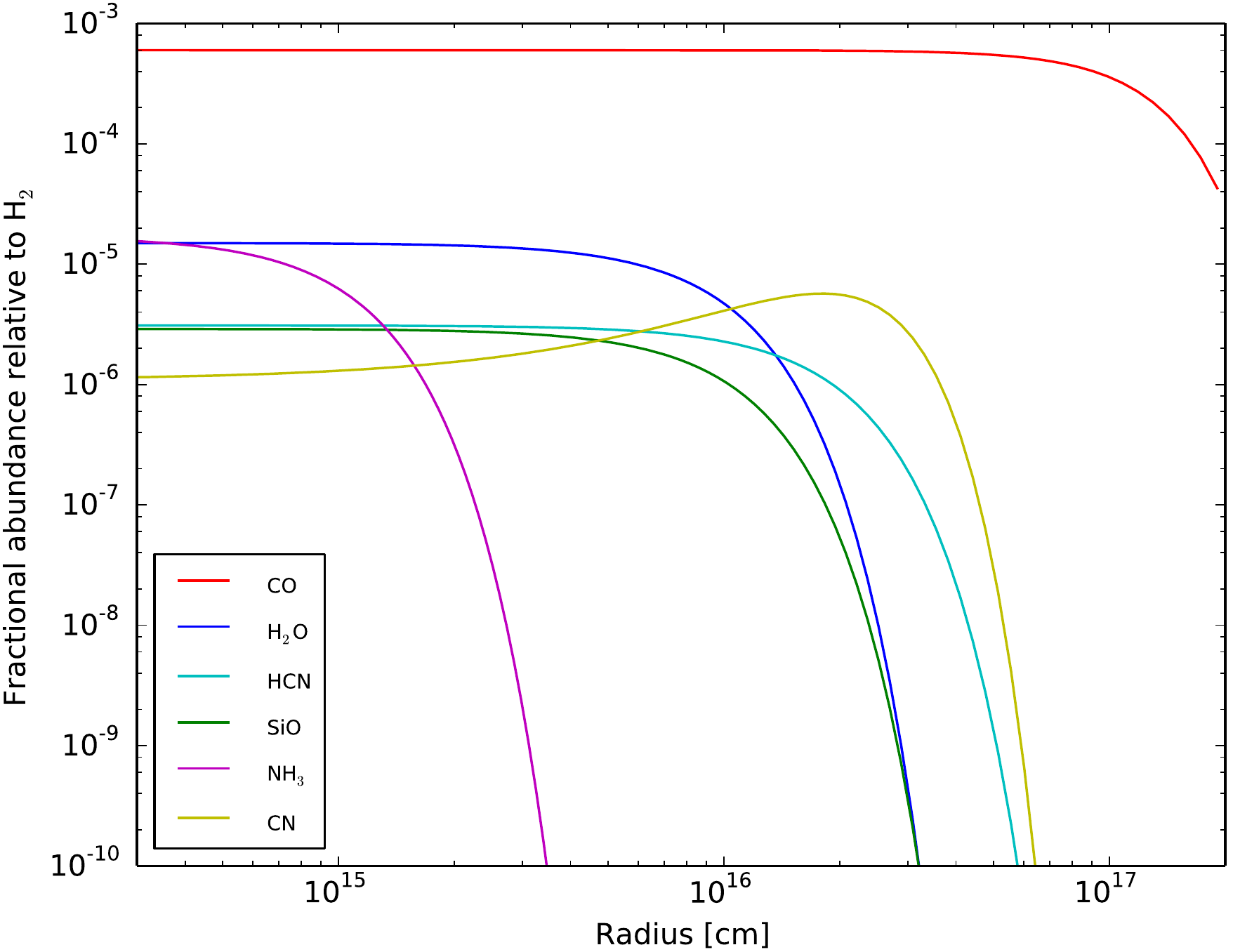}
      \caption{Abundance distributions of the primary isotopologues of the detected molecules.  See Table \ref{molres} for the numerical values.}
         \label{abundances}
   \end{figure}

We have successfully managed to model the intensities of a large number of lines involving states with a large range in excitation energies for a number of molecular species in the circumstellar envelope of W Aql, see Fig. \ref{Jratios}. Nevertheless, there remain issues which are worth discussing.
A summary of the circumstellar molecular abundances is presented in Table \ref{molres}, and a plot of the abundance distributions is shown in Fig. \ref{abundances}. Below we discuss a number of points related to the observational data and the modelling results.

\subsection{Line asymmetries}\label{sec:asymmetry}

The high resolution CO line profiles (ground-based and HIFISTARS) show an excess emission in the blue-shifted part, starting from around $-10$ to $-15\;\kms$ and continuing out to $-20\;\kms$ or further. In particular, the bump can be clearly seen in the CO $J$\,=\,4$\rightarrow$3 and 3$\rightarrow$2 lines, and it appears to be present in the CO $J$\,=\,2$\rightarrow$1 and 1$\rightarrow$0 lines, but is much less pronounced (see Fig. \ref{colines}). In other molecules, it is apparent in the o-\h2O $(1_{1,0}\rightarrow1_{0,1})$ line, and particularly strong in the HCN $J$\,=\,13$\rightarrow$12 line. In HCN this emission cannot be due to hyperfine structure effects as these are small at such high $J$.

A possible explanation for this blue-shifted wing emission is an asymmetric outflow, possibly due to interactions between W Aql and its main sequence companion. \cite{Ramstedt2011} found an asymmetric circumstellar dust feature to the south-west of the star. The dust feature appears to be in the same direction as the companion, but it is unclear if this is significant or merely a projection effect. The main sequence companion could be moving inside the CSE, but W Aql could also be a wider binary, in which case it is unclear if the companion would influence the CSE to that extent. \cite{Mayer2013} imaged the $70\um$ and $160\um$ dust emission in the vicinity of W Aql. They suggest that the shape of the dust distribution around W Aql could be due to a loose Archimedean spiral created by binary interaction. They also calculated the proper motion of W Aql from observations spanning almost a century. Their $\upsilon_\mathrm{LSR}=-21.6\pm 4.1 \;\kms$ agrees with our line-profile value of $-23\;\kms$. 

\subsection{Issues in the line modelling}

A simple dynamical model based on simultaneously solving the velocity equations for the (momentum coupled) dust and the gas was used to derive the gas expansion velocity radial profile (see \citet{Ramstedt2008} for a presentation of the dynamical model). The predicted gas velocity profile is well described by Eq. (2) with $\beta = 0.7$. However, a value of $\beta =2.0$ gives a much better fit to the CO line shapes, where the higher-energy lines are noticeably narrower than the lower-energy lines. Thus, it appears that the gas acceleration is slower than predicted by the dust-driven wind model that we employed. This implies that dust grains in W Aql may not be very efficient in driving the wind and hence are not fully momentum coupled to the gas. Apparently slow rising expansion velocities have also been found in other cases of well-studied AGB stars \citep[see for example][]{Decin2010,Khouri2014}.

The model line shapes of both the $^{12}$CO and $^{13}$CO low-$J$ lines are somewhat double-peaked, indicating that the line emissions are spatially resolved in the model, as opposed to the observed lines that are more rounded. This could be eliminated by either increasing the distance to W Aql, or by decreasing the size of the $^{12}$CO and $^{13}$CO envelopes. A test of the latter only reduced the intensity of the lowest lines with minimal changes to their shape. Changing the distance --- which also requires an increase in mass-loss rate to compensate --- required distances further away than any given in the literature, and even so did not result in a model that consistently fitted the low- and high-$J$ CO lines. The discrepancy in the low-$J$ model lines could also be an effect of imperfections in our circumstellar model, e.g. due to clumpiness or a stronger interstellar radiation field than is used in the calculations, perhaps due to a contribution from the companion. It is difficult, at this point, to firmly resolve this issue, but we note that discrepancies in the form of double-peaked model line shapes and observed single-peaked line shapes are not uncommon \citep{Justtanont2005,Khouri2014,Olofsson2002}. 

{ The combination of lines covering a large range in excitation energies lead to the conclusion that the NH\down{3} envelope is much smaller than all the other molecular envelopes. From a purely photodissociation point of view, NH\down{3} is expected to have a smaller envelope than \h2O as its (unshielded) photodissociation rate is higher \citep[by a factor of $\sim 1.5$,][]{van-Dishoeck2006} and it is relatively less well shielded by interstellar dust. It is unclear whether that difference would be sufficient to reduce the envelope to the size we find for W Aql. There may be a more complicated abundance distribution than assumed here, perhaps due to effects such as condensation onto dust grains. We plan to investigate the distribution of NH\down{3} in other AGB stars to determine whether W Aql is anomalous in this respect, or whether there is an underlying trend.}

In the comparison of model and observed line profiles there are two lines that stick out, the HCN $J$\,=\,3$\rightarrow$2 and SiO $J$\,=\,8$\rightarrow$7 lines are particularly discrepant. We have no reason to believe that these lines are particularly poorly calibrated, and hence other explanations for the discrepancies are warranted. A natural explanation would be time variability, as W Aql is a Mira variable. Detailed observations of the C star IRC+10216 have shown considerable intensity variations in the lines of some circumstellar species (Teyssier et al., in prep). Hence, we cannot exclude variability as an explanation. In the case of SiO another effect may explain the discrepancy. Circumstellar SiO lines are known to be strongly affected by line overlaps \citep{Pardo1998,Herpin2000}. This can lead to peculiar intensities of certain lines, often manifested in the form of maser emission. It is not known whether the same effect can also influence the HCN line, and there is no evidence for this in the study of circumstellar HCN by \citet{Schoier2013}.

Our derived HCN and CN results are not compatible. The derived CN abundance is higher than that of HCN, which is not expected if CN is formed through photodissociation of HCN. We caution that the uncertainties in the abundance estimates are such that the result is marginal. However, the result is in line with the problems that \citet{Schoier2011} had reconciling their HCN and CN observations of $\chi$ Cyg. They were unable to find a CN shell model which agreed with both HIFI and low-$J$ CN data and instead used a Gaussian abundance distribution of the type given in Eq.~\ref{abundance}. In addition, they could not find an HCN model which fitted the HCN $J$\,=\,13$\rightarrow$12 line together with the lower-$J$ lines. { On the contrary, our HCN model holds comparatively well up to the $J$\,=\,28$\rightarrow$27 observed by PACS. Formally, \citet{Schoier2011} derived CN/HCN\,$\approx$\,4, which is even higher than our ratio of two.} It is possible that the behaviour of the HCN and CN molecules in CSEs is not yet fully understood.

\subsection{Comparison with other stars}

Most of the molecules detected in W Aql by HIFI are oxygen-bearing, with HCN the only carbon-bearing molecule seen. The abundances are generally consistent with W Aql being chemically intermediate between an M and a C star, i.e., an S star. { The quoted errors of the abundances are statistical within the adopted circumstellar model. If we include uncertainties in the adopted circumstellar model, in the adopted abundance distributions, etc., the errors in the abundances rise to more than a factor of a few.}

The only other S star observed as part of the HIFISTARS program was $\chi$ Cyg, presented in \citet{Schoier2011}. The two stark differences between W Aql and $\chi$ Cyg in the HIFI observations are the strong CN lines present in $\chi$ Cyg and absent in W Aql, and the NH\down{3} line detected in W Aql but not in $\chi$ Cyg. The two CN lines detected in setting 17 in $\chi$ Cyg are of comparable line intensity to the SiO $J$\,=\,16$\rightarrow$15 line in the same setting and much stronger than the H\up{13}CN $J$\,=\,8$\rightarrow$7 line. For example, the line intensity ratios in $\chi$ Cyg are CN($6_{11/2}\rightarrow 5_{9/2}$)/SiO($16\rightarrow15$) = 1.4 and CN($6_{11/2}\rightarrow 5_{9/2}$)/H\up{13}CN($8\rightarrow7$) = 11, whereas the upper limits for the same ratios in W Aql are $<$\,0.5 and $<$\,0.8, respectively. This implies a greatly reduced CN abundance, at least in the inner CSE, for W Aql relative to $\chi$ Cyg. { On the contrary, the upper limit for the abundance of NH$_3$ is several orders of magnitude lower in $\chi$ Cyg than in W Aql. If, for a closer comparison, we assume the same $e$-folding radius in $\chi$ Cyg as in W Aql, we find that the upper limit of the NH\down{3} abundance in $\chi$ Cyg is a factor of 30 lower than the abundance in W Aql.}

Other than CN and NH\down{3}, the circumstellar abundances determined for W Aql in this paper are mostly comparable to those \citet{Schoier2011} derived for $\chi$ Cyg; there is slightly more \h2O, 1.6 times as much HCN, and about a factor of 5 more SiO in $\chi$ Cyg. The difference in HCN and CN between W Aql and $\chi$ Cyg is surprising as, generally, both molecules are thought to be more abundant in more carbon-rich stars. Optical spectral classification puts W Aql on the carbon end of the S star scale with a high C/O ratio, one step from an SC classification, while $\chi$ Cyg has a lower C/O ratio and is close to being an MS star \citep{Keenan1980}. The higher abundance of SiO in $\chi$ Cyg, on the other hand, is in agreement with the respective C/O ratios of the two S stars.

The abundance of \h2O in W Aql lies between the abundances generally derived for M stars, $\sim10^{-4}$ \citep{Maercker2008}, and the much lower abundances being found for C stars \citep{Neufeld2011}. The \h2O ortho/para ratio is in line with the expected value of 3 for \h2O formation in warm thermal equilibrium conditions.

\begin{figure}[tb]
   \centering
   \includegraphics[width=7cm]{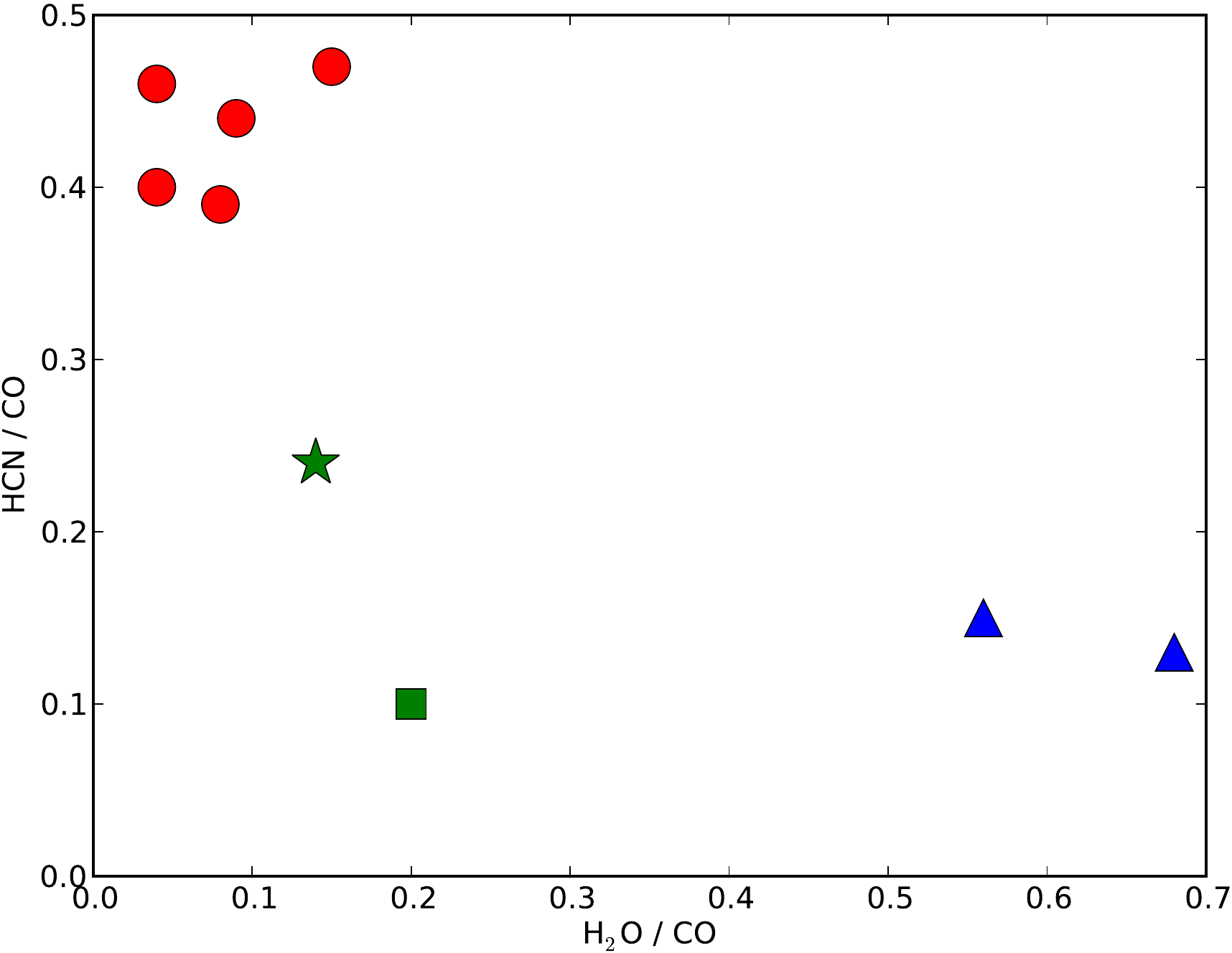}
   \includegraphics[width=7cm]{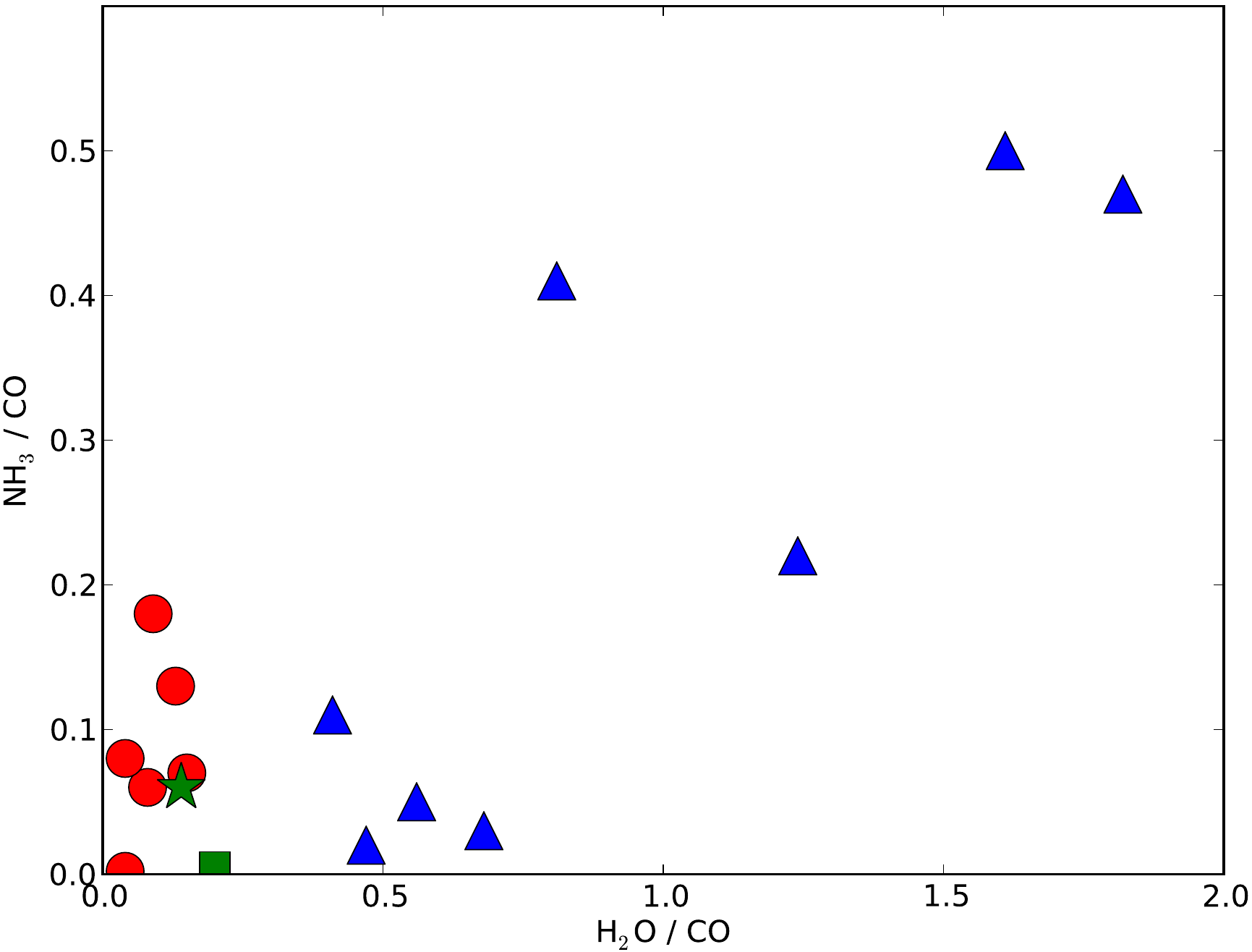}
      \caption{Line intensity ratios of HCN($13\rightarrow12$)/CO($6\rightarrow5$) and \h2O($(1_{1,0}\rightarrow1_{0,1})$)/CO($6\rightarrow5$) plotted against each other {(top)}, and of NH\down{3}($1_0\rightarrow0_0$)/CO($6\rightarrow5$) and \h2O($(1_{1,0}\rightarrow1_{0,1})$)/CO($6\rightarrow5$) plotted against each other {(bottom)}. The data are given in Table \ref{CMratios}. The red circles are C stars, the blue triangles are M stars, and the two S stars are shown in green with W Aql being the star and $\chi$ Cyg the square. In the lower plot, the ratio shown for NH\down{3}/CO for $\chi$ Cyg is an upper limit. The $\chi$ Cyg values are taken from \citet{Schoier2011}.}
         \label{ratioplots}
   \end{figure}

{ In order to further investigate possible trends in the chemistry, we analysed a number of line-intensity ratios. This has the advantage of being model-independent, but requires that the locations of the individual molecular species and the excitation of the individual lines used are similar in all objects to be able to draw conclusions. We believe that the chosen lines (NH\down{3}($1_0 \rightarrow 0_0$), HCN($13\rightarrow12$), o-\h2O $(1_{1,0}\rightarrow1_{0,1})$, and CO($6\rightarrow5$)) fulfil these criteria.} We compare these line intensities with the corresponding results for the M stars in the HIFISTARS sample presented in \cite{Justtanont2012} and the C stars from the HIFISTARS sample where both \h2O and NH\down{3} lines have been detected (M Schmidt, private correspondence). For each star, we normalised the line intensities with the CO($6\rightarrow5$) line intensity. The relevant line intensities and ratios are given in Table \ref{CMratios}. The relationships between  the \h2O and NH\down{3} lines, and the \h2O and HCN lines are plotted in Fig. \ref{ratioplots}.

Comparing the trends depicted in Fig. \ref{ratioplots} there is a clear demarcation between M and C stars in the HCN/CO against \h2O/CO diagram. The two S stars lie between the M and C areas of the plot, a result consistent with their S-type classification. In terms of HCN, the S stars exhibit more M star-like properties, and in terms of \h2O they exhibit more C star-like properties. Examining the NH\down{3}/CO against \h2O/CO diagram, we see a different relationship. There appears to be a strong positive correlation between NH\down{3} and \h2O in M stars, while for C stars the \h2O line flux seems to be independent of the NH\down{3} line flux. The S stars seem to fall in the overlapping region, perhaps with W Aql behaving more like a C star, as might be expected given its spectral classification. 

\begin{table*}[tb]
\caption{NH\down{3}($1_0\rightarrow0_0$), CO($6\rightarrow5$), o-\h2O($1_{1,0}\rightarrow1_{0,1}$) and HCN($13\rightarrow12$) line intensity ratios for W Aql and the M and C stars in HIFISTARS for which NH\down{3} was detected (except IRC+10216).}
\label{CMratios}      
\centering
\begin{tabular}{llcccccccc}
\hline\hline
Classification & Star & $I_{\mathrm{NH}_3}$ & $I_\mathrm{CO}$ & $I_{\mathrm{o-H}_2\mathrm{O}}$& $I_\mathrm{HCN}$ & $I_{\mathrm{NH}_3}/I_\mathrm{CO}$ & $I_{\mathrm{o-H}_2\mathrm{O}}/I_\mathrm{CO}$ & $I_{\mathrm{HCN}}/I_\mathrm{CO}$ & Ref.\\
&&[K $\kms$]&[K $\kms$]&[K $\kms$]&[K $\kms$]\\
\hline
S stars & W Aql		& $\;$0.85	& 14.1\phantom{0}	& 2.0 & 3.5\phantom{0} & 0.06 & 0.14 & 0.25 & This work\\
		& $\chi$ Cyg& $<0.07\;\;$ & 14.3\phantom{0}& \phantom{0}2.91 & 1.49& $<0.004\;\,$ & 0.20 & 0.10 & 3\\
\hline
M stars & IRC+10011	&	2.2	& 4.4 	& 7.1 &...& 0.50 & 1.61&...&1\\
&IK Tau		&	4.7	& 11.6\phantom{0}	& 9.4 &...& 0.41 & 0.81&...&1\\
&R Dor		&	0.5	& 14.9\phantom{0}	& 10.2\phantom{0} &1.9& 0.03 & 0.68&0.13&1\\
&TX Cam		&	1.5	& 14.2\phantom{0}	& 5.8 &...& 0.11 & 0.41&...&1\\
&W Hya		&	0.5	& 9.6	& 5.4 &1.4& 0.05 & 0.56&0.15&1\\
&AFGL 5379	&	1.7	& 7.8	& 9.7 &...& 0.22 & 1.24&...&1\\
&OH 26.5		&	0.8	& 1.7	& 3.1 &...& 0.47 & 1.82&...&1\\
&R Cas		&	0.3	& 14.7\phantom{0}	& 6.9 &...& 0.02 & 0.47&...&1\\
\hline
C stars &V Hya		& $\;\,$0.03	& 13.9	& \phantom{0}0.6	& \phantom{0}5.6& $\;\,$0.002& 0.04 & 0.4& 2\\
&RW LMi		& $\;\,$2.1\phantom{0} & 34.1 & \phantom{0}2.9 & 13.3 & 0.06 & 0.08 & 0.4 & 2\\
&II Lup		& $\;\,$3.0\phantom{0} & 16.4 & \phantom{0}1.4 & \phantom{0}7.2 & 0.2\phantom{0} &0.09 & 0.4 & 2\\
&LP And		& $\;\,$0.7\phantom{0} & \phantom{0}9.0 & \phantom{0}0.4 & \phantom{0}4.2 &0.08 &0.04 &0.5 & 2\\
&V Cyg		& $\;\,$0.6\phantom{0}	& \phantom{0}9.1 & \phantom{0}1.4	& \phantom{0}4.3 & 0.07 & 0.15  & 0.5 & 2\\
&V384 Per	& $\;\,$0.8\phantom{0}	& 	\phantom{0}6.0 & \phantom{0}0.8	& ...& 0.13 & 0.13 & ...& 2\\
\hline
\end{tabular}
\tablefoot{References: (1) intensity values taken from \cite{Justtanont2012}; (2) intensities from M Schmidt, private correspondence; (3) intensities from \citet{Schoier2011}.}
\end{table*}

\section{Conclusions}

The main conclusions of this paper are as follows:

   \begin{itemize}
      \item The HIFI spectrum observed for the AGB star W Aql contains spectral lines from a variety of circumstellar molecules, CO, SiO \h2O, HCN, and the first detection of NH\down{3} in an S star.
      \item Excess emission was detected on the blue side of several molecular lines, most noticeably in the HIFI detections of CO lines, HCN lines, and the ground-state o-\h2O line. We surmise that this could be due to an asymmetric outflow.
      \item { We have successfully performed a radiative transfer analysis of the ground-based, HIFI, and PACS observations of the detected molecules, and we were able to include \h2O line cooling in the gas energy balance equation.}
      \item The acceleration of the gas appears to be slow compared to that predicted by a dust-driven wind model.
      \item The mass-loss rate of W Aql is estimated to be \mbox{$4.0\e{-6} \spy$}, and the gas expansion velocity is $16.5\;\kms$.
      \item The estimated abundances of circumstellar SiO, \h2O and HCN clearly place W Aql in between the M and C stars chemically, i.e., the circumstellar abundances are consistent with an S star classification.
      \item The CSE of W Aql contains approximately a fifth of the SiO present in the other S star examined in detail in the HIFISTARS project, $\chi$ Cyg, but this is consistent with the spectral classification of W Aql, which puts it on the more carbon-rich end of the S star scale. On the other hand, it contains less CN and HCN than $\chi$ Cyg, which is unexpected.
      \item { By combining PACS and HIFI data, we were able to constrain both the abundance and the $e$-folding radius of the NH\down{3} envelope. We found a much smaller envelope than previously assumed and a relatively high abundance.}
      \item An analysis of line intensity ratios, based on specific CO, \h2O, HCN and NH\down{3} lines among AGB stars confirms that W Aql lies chemically between C stars and M stars. In terms of \h2O, W Aql is more similar to the C stars and in terms of HCN, W Aql is more similar to the M stars.
      \item It has not been possible to find a model that consistently fits both the HCN and CN data.
      \item In our CO analysis, we found a \up{12}CO/\up{13}CO ratio of 29, in line with expectations for S stars.
   \end{itemize}

\begin{acknowledgements}
		The authors would like to acknowledge Lewis Hutton, whose programming expertise was invaluable in the preparation of this paper, John Black, for his compilation of the NH\down{3} molecular data file, and Rafael Bachiller, for providing us with the ground-based CN data. The authors are also grateful to the referee for providing many constructive comments. TD and KJ acknowledge funding from the SNSB.
		
		HIFI has been designed and built by a consortium of institutes and university departments from across Europe, Canada and the United States under the leadership of SRON Netherlands Institute for Space Research, Groningen, The Netherlands and with major contributions from Germany, France and the US. Consortium members are: Canada: CSA, U.Waterloo; France: CESR, LAB, LERMA, IRAM; Germany: KOSMA, MPIfR, MPS; Ireland, NUI Maynooth; Italy: ASI, IFSI-INAF, Osservatorio Astrofisico di Arcetri-INAF; Netherlands: SRON, TUD; Poland: CAMK, CBK; Spain: Observatorio Astron\'omico Nacional (IGN), Centro de Astrobiolog\'ia (CSIC-INTA). Sweden: Chalmers University of Technology - MC2, RSS \& GARD; Onsala Space Observatory; Swedish National Space Board, Stockholm University - Stockholm Observatory; Switzerland: ETH Zurich, FHNW; USA: Caltech, JPL, NHSC.
		
		PACS has been developed by a consortium of institutes led by MPE (Germany) and including UVIE (Austria); KU Leuven, CSL, IMEC (Belgium); CEA, LAM (France); MPIA (Germany); INAF-IFSI/OAA/OAP/OAT, LENS, SISSA (Italy); IAC (Spain). This development has been supported by the funding agencies BMVIT (Austria), ESA-PRODEX (Belgium), CEA/CNES (France), DLR (Germany), ASI/INAF (Italy), and CICYT/MCYT (Spain).
      
      Based on data obtained from the ESO Science Archive Facility under request numbers 47181, 45037, 46391, 53323, 45256 and 53506.
      
      This publication is based on data acquired with the Atacama Pathfinder Experiment (APEX). APEX is a collaboration between the Max-Planck-Institut f\"ur Radioastronomie (Germany), the European Southern Observatory, and the Onsala Space Observatory (Sweden).
      
      This publication makes use of data products from the Two Micron All Sky Survey, which is a joint project of the University of Massachusetts and the Infrared Processing and Analysis Center/California Institute of Technology, funded by the National Aeronautics and Space Administration and the National Science Foundation.

\end{acknowledgements}

\bibliographystyle{aa}
\bibliography{WAql_3.bbl}

\appendix
\section{Unidentified PACS lines}\label{usec}

 When identifying lines to include in our modelling, we excluded lines which were not clear single-transition identifications. Primarily, this involved excluding lines where the FWHM of the fitted Gaussian profile was larger than the PACS spectral resolution by at least 20\% (the FWHM criterion) and lines with double identifications, especially for CO and \h2O blends. However, we retained some lines in our modelling which we identified as CO or \h2O plus a significantly weaker line. These were flagged in Table \ref{pacslines}.

In Table \ref{Ulines} we collect all the unidentified PACS lines as well as all the blended lines. Where we have made a partial identification, that is also listed. Lines which are blended based on the FWHM criterion are labelled as such. Lines which partially overlap --- that is, there are two (or more) distinct peaks but the line partially overlaps with the neighbouring line(s) --- are also labelled. The three blended lines that we retained in our modelling are also clearly labelled, with the primary identification listed first and the weaker blended line in brackets. The bands listed in Table \ref{Ulines} are as described in Sect \ref{obs:pacs}. All the wavelengths listed are observed wavelengths, not corrected for the Doppler shift due to the stellar velocity.

\begin{table*}[tb]
\caption{Unidentified and blended PACS detections}
\label{Ulines}
\centering         
\begin{tabular}{ccccl} 
\hline\hline  
Peak $\lambda$	&	Band	&	Flux	&	Flux error	&		Notes	\\
$[\mu$m$]$	&		&	[W m$^{-2}$]	&	[W m$^{-2}$]	&			\\
\hline										
179.519	&	R1B	&	2.0E-16	&	8.3E-18	&			\\
174.605	&	R1B	&	1.7E-16	&	9.1E-18	&			\\
170.255	&	R1B	&	3.8E-17	&	9.6E-18	&		FWHM blend, overlap	\\
165.741	&	R1B	&	5.0E-17	&	1.1E-17	&	NH\down{3} + ?, overlap	\\
165.606	&	R1B	&	7.1E-17	&	1.4E-17	&		overlap	\\
165.187	&	R1B	&	3.0E-17	&	7.6E-18	&			\\
156.216	&	R1B	&	1.4E-16	&	1.1E-17	&			\\
144.782	&	R1B	&	4.7E-16	&	9.9E-18	&			\\
144.510	&	R1B	&	9.1E-17	&	1.1E-17	&			\\
143.493	&	R1B	&	4.8E-17	&	8.3E-18	&			\\
144.519	&	R1A	&	9.6E-17	&	1.0E-17	&			\\
143.496	&	R1A	&	3.5E-17	&	3.8E-18	&			\\
141.277	&	R1A	&	6.8E-17	&	7.7E-18	&			\\
136.511	&	R1A	&	1.5E-16	&	1.2E-17	&		FWHM blend, overlap	\\
136.344	&	R1A	&	5.8E-17	&	8.8E-18	&		overlap	\\
130.387	&	R1A	&	5.2E-16	&	1.5E-17	&	CO (+ HCN), included	\\
124.912	&	R1A	&	2.6E-16	&	1.5E-17	&		FWHM blend, overlap	\\
124.679	&	R1A	&	1.4E-16	&	1.6E-17	&	NH\down{3} + ?,	FWHM blend, overlap	\\
113.500	&	R1A	&	7.7E-16	&	2.2E-17	&	CO + \h2O,	FWHM blend	\\
103.920	&	R1A	&	8.4E-17	&	2.1E-17	&			\\
~~94.707	&	B2B	&	1.0E-16	&	1.2E-17	&		overlap	\\
~~94.644	&	B2B	&	1.1E-16	&	1.4E-17	&		FWHM blend, overlap	\\
~~92.810	&	B2B	&	6.3E-17	&	1.2E-17	&		FWHM blend	\\
~~87.186	&	B2B	&	2.8E-16	&	1.9E-17	&	CO + ?,	FWHM blend	\\
~~84.767	&	B2B	&	1.4E-16	&	1.6E-17	&	NH\down{3} + ?,	FWHM blend, overlap	\\
~~84.692	&	B2B	&	1.3E-16	&	2.7E-17	&		FWHM blend, overlap	\\
~~84.619	&	B2B	&	7.3E-17	&	1.6E-17	&		overlap	\\
~~84.536	&	B2B	&	1.2E-16	&	1.8E-17	&		FWHM blend	\\
~~84.410	&	B2B	&	2.9E-16	&	1.2E-17	&			\\
~~83.842	&	B2B	&	1.6E-16	&	3.0E-17	&		FWHM blend	\\
~~83.279	&	B2B	&	1.5E-16	&	1.9E-17	&			\\
~~82.978	&	B2B	&	8.3E-17	&	2.0E-17	&			\\
~~78.933	&	B2B	&	1.1E-16	&	2.0E-17	&			\\
~~75.911	&	B2B	&	6.0E-17	&	2.0E-17	&			\\
~~75.389	&	B2B	&	5.4E-16	&	2.3E-17	&	\h2O (+ \h2O), included	\\
~~74.948	&	B2B	&	1.2E-16	&	2.0E-17	&		overlap	\\
~~74.891	&	B2B	&	1.8E-16	&	2.5E-17	&	CO + ?,	FWHM blend, overlap	\\
~~72.439	&	B2B	&	8.3E-17	&	1.9E-17	&			\\
~~71.967	&	B2B	&	3.6E-16	&	3.7E-17	&		FWHM blend	\\
~~71.606	&	B2B	&	9.9E-17	&	2.9E-17	&		overlap	\\
~~71.543	&	B2B	&	1.3E-16	&	2.5E-17	&		overlap	\\
~~71.079	&	B2B	&	2.8E-16	&	3.1E-17	&		FWHM blend	\\
~~72.839	&	B2A	&	9.3E-17	&	1.9E-17	&			\\
~~72.517	&	B2A	&	6.9E-17	&	1.7E-17	&			\\
~~72.435	&	B2A	&	9.3E-17	&	1.8E-17	&			\\
~~71.955	&	B2A	&	3.1E-16	&	2.3E-17	&		FWHM blend	\\
~~71.605	&	B2A	&	1.2E-16	&	2.2E-17	&	NH\down{3} + ?, overlap	\\
~~71.534	&	B2A	&	1.8E-16	&	2.1E-17	&		overlap	\\
~~71.070	&	B2A	&	2.4E-16	&	2.1E-17	&			\\
~~70.903	&	B2A	&	5.6E-17	&	1.7E-17	&			\\
~~70.711	&	B2A	&	9.1E-17	&	1.8E-17	&			\\
~~70.287	&	B2A	&	7.5E-17	&	2.3E-17	&		FWHM blend	\\
~~69.075	&	B2A	&	1.0E-16	&	2.6E-17	&	CO + ?,	FWHM blend	\\
~~68.483	&	B2A	&	7.1E-17	&	1.3E-17	&			\\
~~67.361	&	B2A	&	1.0E-16	&	2.5E-17	&	CO, overlap	\\
~~67.283	&	B2A	&	2.8E-16	&	2.9E-17	&		FWHM blend, overlap	\\
~~67.095	&	B2A	&	3.6E-16	&	2.3E-17	&	\h2O (+ NH\down{3}), included	\\
~~66.104	&	B2A	&	2.5E-16	&	1.7E-17	&			\\
~~65.175	&	B2A	&	3.1E-16	&	2.1E-17	&			\\
~~63.468	&	B2A	&	2.1E-16	&	4.1E-17	&		FWHM blend	\\
~~63.325	&	B2A	&	3.3E-16	&	3.8E-17	&	NH\down{3},	FWHM blend	\\
~~60.504	&	B2A	&	1.6E-16	&	2.4E-17	&			\\
~~60.177	&	B2A	&	8.0E-17	&	1.7E-17	&	\h2O blend		\\
~~58.716	&	B2A	&	5.0E-16	&	2.1E-17	&			\\
~~57.648	&	B2A	&	3.4E-16	&	3.3E-17	&			\\
~~56.806	&	B2A	&	3.5E-16	&	4.7E-17	&		FWHM blend	\\
\hline
\end{tabular}
\tablefoot{See text in Appendix \ref{usec} for full explanation.}
\end{table*}

\end{document}